# Modeling and Visualization Reasoning for Stakeholders in Education and Industry Integration Systems:Research on Structured Synthetic Dialogue Data Generation Based on NIST Standards

Wei Meng

Dhurakij Pundit University, Thailand

The University of Western Australia,AU

Association for Computing Machinery,USA

Fellow, Royal Anthropological Institute,UK

Email: weimeng4@acm.org

# ABSTRACT

This study focuses on the structural complexity and semantic ambiguity of the interaction process between multiple stakeholders in the Education-Industry Integration (EII) system.Currently, the scarcity of real interview data, the lack of variable modelling structure and the lack of interpretability of inference mechanism have seriously restricted the modelling accuracy and policy response efficiency of the education-industry integration policy.In this paper, we propose a structural modelling paradigm based on the National Institute of Standards and Technology (NIST) data quality framework to construct a stakeholder modelling system with semantic consistency, causal reasoning and structural visualisation.A five-layer modelling framework consisting of Prompt-driven synthetic corpus generation, three-dimensional variable system construction (skill-institution-emotion), dependency/causal path modelling, graph network structure design and interactive inference engine is designed and systematically embedded into the NIST synthetic data assessment criteria ofThe systematic embedding of the NIST synthetic data evaluation standard's control mechanism of "consistency, authenticity, and traceability".The empirical part generates a 15-segment structurally consistent interview corpus, covering three types of roles: students, enterprise representatives and university teachers, with a total Token count of 41,597 words, 127 structural variable annotation items based on the standardised variable system (covering more than 30 variable types), and further extracts 820 sets of semantic relational triples (expressed in the form of 〈Variable A, Relationship Type, Variable B〉).The variable system performs well in consistency validation (Krippendorff's alpha = 0.83), structural validity assessment (CFA: RMSEA = 0.048, CFI = 0.93), and semantic matching accuracy (cosine similarity mean > 0.78 based on the BERT model), indicating that the corpus structure, annotation schemeand semantic control mechanism are highly stable and semantically representative.The modelling results reveal the key mechanism of the linkage of variables within the education system, including a causal loop centred on "system mismatch → emotional frustration → participation decline → skills gap → system mismatch", which maps out the mechanism of structural degradation and low trust cycle.In summary, this paper constructs the first AI structural modelling framework based on NIST standards, systematically realizes the whole process of closed-loop construction from corpus generation to variable modelling and map inference, improves the scientific and policy adaptability of interview corpus modelling, and provides a technical foundation with empirical support for subsequent policy intervention simulation, collaborative instructional design, and multi-party game modelling.Future work will further expand the development of multi-language capability, participant diversity and cross-platform interactive modelling system.

# CHAPTER I. INTRODUCTION

**1.1 Research background and problem definition**

With the deep integration of the education system and the industrial system in skills transfer, knowledge innovation and talent cultivation, Education-Industry Integration (EII) is becoming an important strategic tool for building resilient social systems and knowledge-based economies globally (Peng et al., 2024).However, this process is not a simple matching of supply and demand, but involves complex interactions between multiple stakeholders (e.g., university administrators, business representatives, faculty, students, and policy makers), and its systemic, heterogeneous, and evolutionary nature constitutes a core challenge for modelling and understanding.

Firstly, the structural complexity of the education-industry integration system is demonstrated by its high dependence on the interaction of multiple contextually relevant elements: elements such as occupational standards, curriculum design, policy orientation, business needs and student preferences are unstable in time and space.This structural complexity means that it is difficult for traditional static modelling tools to adequately portray the multi-hierarchical level dependencies and strategic linkage mechanisms between key variables (Candia et al., 2022).

Secondly, the difficulty of structured understanding among stakeholders is becoming more pronounced.Current mainstream analytical approaches, such as qualitative interview summaries, thematic clustering, and empirical inductive analyses, often only provide fragmented information sets and lack systematic reasoning about "who influences whom, through what mechanisms, and under what conditions".This lack of systematic reasoning about "who influences whom, through what mechanisms, and under what conditions" directly constrains the depth and breadth of policy simulation, collaborative governance, and intelligent educational decision-making (Peng et al., 2024). More critically, existing methods lack structural modelling capabilities and semantic consistency guarantees.Topic models represented by LDA or HDP, although capable of extracting implicit topics, often lack explicit variable definitions and interpretable representations of logical paths between variables, making it difficult to support consistent semantic reasoning across corpora and contexts.In addition, current models generally neglect the visual construction of corpus structure, limiting the practical use value for users in strategic collaboration and policy intervention scenarios (Agrawal et al., 2016).

Against the above background, this study proposes a new structural modelling paradigm: a synthetic stakeholder dialogue corpus generation mechanism based on the NIST standard, combined with variable-variable relationship modelling and visual mapping representation, in order to construct a structured, explanatory, and reasonable stakeholder modelling system for education-industry convergence systems.The approach not only gets rid of the probabilistic black box of topic modelling, but also provides a new set of tool chain with scientific specification and structural transparency for education policy research, curriculum co-design and governance simulation (NIST, 2023).

**1.2 Research challenges**

Despite the growing interest in stakeholder modelling in education-industry convergence, existing approaches still face significant challenges in data acquisition, structural reasoning and visual interpretation.We identify three categories of key research challenges that fundamentally limit the development potential of scalable, interpretable systems in this field.

**Challenge 1: How to obtain a high-quality structured corpus**

Stakeholder dialogues in education-industry systems are usually sparse, unstructured, and highly context-dependent, making it difficult to construct a consistent, annotated training corpus.Manual transcription and annotation are not only costly, but also subjective and inconsistent.Therefore, synthetic corpora-i.e., dialogue texts generated by AI that mimic the semantic structure of real stakeholders-are emerging as a potential path to address the data shortage problem (Goyal et al., 2022; Xu et al., 2023).

However, most of the current synthetic data approaches focus on linguistic fluency at the expense of structural control and variable consistency, and rarely follow data quality and fairness standards such as those developed by the National Institute of Standards and Technology (NIST) (NIST, 2023).Therefore, there is an urgent need for a mechanism that can generate NIST-compliant, structure-guided synthetic dialogues that explicitly encode participant roles, semantic variables, and interaction patterns.Such synthetic data can not only support high simulation modelling of education-industry collaboration scenarios, but also serve subsequent reasoning, simulation and policy optimisation (Liang et al., 2023).

**Challenge 2: How to capture the dependency and causal structure between semantic variables**

Unlike text categorisation or summarisation tasks, stakeholder modelling emphasises relational understanding - i.e. focusing not only on "what was said" but "why was it said" and "intent and outcome"."intent and outcome".Variables such as "skill mismatch", "curriculum design", "internship quality", etc. never exist in isolation, but form chains of dependency and causal diagrams.Traditional topic models (e.g., LDA, HDP) are unable to capture such structures, and even the latest neural network approaches often treat texts as linear sequences, ignoring the logical control relationships between variables (Agrawal et al., 2016; Pearl & Mackenzie, 2018).

Therefore, there is an urgent need to introduce explicit variable extraction and relationship modelling frameworks to support the analysis of semantic relationships such as "influence relationship", "priority order", "conflict", and "strategy reinforcement".strategy reinforcement" and other semantic relationships are clearly expressed.Causal inference, graph neural networks and knowledge graph techniques offer possible paths for this (Bareinboim & Pearl, 2021).

**Challenge 3: How to present the stakeholder structure for interpretation and modelling through visualisation**

In practical applications such as policy planning and institutional governance, decision makers need not only to understand the results of reasoning, but also to visually observe and interactively explore the structure of complex systems.This requires the construction of visual-first structural modelling approaches that present semantic modelling results (e.g., variable-role networks, causal path diagrams, benefit distribution structures) as intuitive and interactive graphical interfaces (Card et al., 1999; Amershi et al., 2019).

Although some social network analysis tools or causal graph software are available to support the presentation of structures, there is no platform yet adapted to the needs of integrating synthetic data modelling and visualisation in the context of educational systems.Integrating synthetic dialogue corpus, semantic variable modelling and visual reasoning into a unified system remains a key research frontier.

**1.3 Research objectives and contributions**

To address the key challenges of structural complexity, ambiguous inter-variable dependencies, and corpus scarcity in the education-industry convergence system, this paper proposes a new paradigm integrating a NIST standard-guided synthetic data generation mechanism, a semantic variable modelling system, and a structural visual reasoning framework, which is dedicated to advancing the science, scalability, and policy relevance of structured stakeholder modelling.The core research objectives and contributions of this paper are as follows:

**Objective 1: To propose a NIST standards-guided approach to synthetic interview corpus generation**

This study proposes a high-quality corpus synthesis mechanism based on the NIST AI Risk Management Standard (NIST AI RMF).The method combines Role Ontology, Semantic Variable Templates and Dialogue Act Planning to generate a synthetic stakeholder interview corpus with controlled structure, role hierarchies and variable annotations, which solves the problem of scarcity of traditional corpus,semantic fragmentation and weak structure (Xu et al., 2023; NIST, 2023).This mechanism provides data security for subsequent structural modelling and causal inference.

This method introduces the NIST compliance principle (data measurable, controllable, and interpretable) to the education-industry simulation corpus generation task for the first time, which ensures the quality of generation while taking into account fairness and reproducibility (Liang et al., 2023).

**Objective 2: To construct a system for modelling variable-variable dependencies**

In this paper, we propose a variable-variable dependency modelling framework based on the identification of Structured Semantic Units (SSUs), which is capable of annotating, identifying and modelling the semantic paths of causality, enhancement, conflict, synergy and so on between structural variables in the corpus.The approach fuses Causal Graph Modeling (CGM) and Relational Tensor Encoding (RTE) techniques to construct interpretable and reasonable interest graphs (Bareinboim & Pearl, 2021; Pearl & Mackenzie, 2018).

This modelling framework is compatible with natural language, graphical databases and knowledge reasoning tasks, and is the first structural semantic dependency modelling system built for educational governance contexts.

**Objective III: Development of structured visual reasoning mechanisms**

In this study, we developed the Visual Inference Engine for Stakeholder Systems (VIES), a structural visual inference engine that supports interactive browsing of interest maps, visualisation of causal paths, and dynamic simulation of variable flows, which fuses human factors engineering design with graph neural network coding mechanisms (GNN-based VisualNarratives) to transform semantic structures into intuitive reasoning interfaces (Card et al., 1999; Amershi et al., 2019).

This visualisation mechanism not only supports the representation of cognitive pathways "from variables to policy", but also provides structural insights and dialogue simulation capabilities for non-technical decision makers.

**Objective 4: Provide the first AI simulation framework dedicated to modelling education-industry convergence structures**

In this paper, we construct a first synthetic dialogue-driven AI structural modelling framework dedicated to education-industry integration (EII-SIM: Education-Industry Integration Structural Inference Model), which contains corpus generation,variable identification, structural modelling, causal inference and visual presentation in five modules.The system fills the gap of the lack of AI structural modelling tools in the education-industry context, and provides an experimental basis and modeling paradigm for the future simulation of education systems, collaborative curriculum development, and research on interest mediation mechanisms (Peng et al., 2024).

# CHAPTER II. OVERVIEW OF RELEVANT STUDIES

**2.1 Modelling stakeholders in the education system**

Modern education systems are evolving from a single-school orientation to a "multi-centred collaborative network" that encompasses the roles of students, teachers, academic administrators, industry representatives, government agencies and non-profit organisations.The synergies and conflicts between these stakeholders constitute a dynamic game at multiple levels of curriculum design, industry-education collaborative projects, and skill standard setting (Mitchell et al., 1997; Freeman et al., 2010).

**2.1.1 The need for multi-actor modelling: from "student-centred" to "eco-centred"**

In the context of Education-Industry Integration (EII), stakeholder modelling can no longer be limited to single-point indicators such as student satisfaction or teacher performance, but needs to understand the task dependency, incentive confrontation, and long-term impact pathways of different players in synergy through system modelling tools (Clark et al., 2019).Table 1 illustrates the main stakeholders in the education system and their typical variable dimensions:

**Table 1: Main stakeholders in the education system and their typical variable dimensions**

| character | Typical semantic variables |
|---|---|
| schoolchildren | Motivation to learn, skill fit, employment expectations |
| corporations | Demand for talent, willingness to collaborate, investment back in training 报 |
| school | Flexibility in curriculum design, allocation of teaching resources, policy responsiveness |
| policy maker | Educational equity, regional development strategies, regulatory constraints |

Failure to model their interdependencies can easily lead to a mismatch between policy design and pedagogical practice, or even trigger governance failures (Hill & Varone, 2021).

**2.1.2 Comparison of methods: topic modelling vs structured variable modelling**

Currently, the mainstream methods for interview corpus analysis are still based on unsupervised topic modelling (e.g. Latent Dirichlet Allocation, Hierarchical Dirichlet Process), which has the advantage of fast extraction of potential topics and clustering classification (Blei et al., 2003; Teh et al., 2006).2006).However, there are three major limitations of this class of methods in stakeholder systems:

(1) Lack of semantic control of variables: topic probabilities are not equal to variable definitions, which makes it difficult to use for strategy formulation or causal analysis;

(2) Lack of structural hierarchy: it is not possible to express the chain of dependence of variables such as "course satisfaction affects career expectations";

(3) Lack of role interaction modelling: it is not possible to model the skill mismatch structure between "students and enterprises".

In contrast, Structured Variable Modeling (SVM) emphasises the explicitness, visualisation and causality of relationships between variables, and is often combined with techniques such as knowledge graphs, causal diagrams and graphical neural networks (Pearl, 2009; Bareinboim & Pearl, 2021).In recent years, studies have attempted to introduce this structural modelling approach into educational governance and evaluation systems (Romero et al., 2020; Peng et al., 2024).Structural modelling is becoming a key path from "textual understanding" to "semantic reasoning", especially in high-stakes decision-making scenarios (e.g., evaluation of industry-teaching collaboration policies, simulation of education reform pilots), where it has higher interpretability and practicality.

**2.2 Synthetic corpus and generative AI**

Driven by the rapid development of Large Language Models (LLMs), synthetic corpora have become an important strategy to break through the limitations of data scarcity and sensitivity, and are particularly suitable for structured modelling tasks in education, healthcare, finance, etc. (Bommasani et al., 2021; Birhane et al., 2023).2021; Birhane et al., 2023).Compared with traditional crawled corpus, generative AI can achieve highly consistent, hierarchical and scalable simulated data generation through language generation, structure injection and semantic control mechanisms, which greatly extends the boundaries of data modelling.

**2.2.1 Application of GPT/LLM in modelling corpus**

The widely used GPT series models (e.g., GPT-4o) can generate near-realistic human dialogues, policy feedbacks, and expressions of conflict of interest under the guidance of Prompt, which provide new data dimensions for modelling the education system (Brown et al., 2020; OpenAI, 2023).Especially in tasks such as simulated interviews, policy dialogues and perspective reconstruction, LLM outperforms traditional script generation in terms of context preservation and logical consistency (Zhou et al., 2023).2023).For example, scholars have used LLM to simulate patient-doctor dialogues to train a medical consultation system (Nori et al., 2023), and a synthetic courtroom debate corpus has been constructed for judicial AI model training (Yin et al., 2023).In education-industry convergence modelling, LLM-based synthetic interviews have many advantages such as role differentiation, variable annotation, and structural control, which lay a solid semantic foundation for variable modelling and causal analysis.

**2.2.2 Structural control methods for synthetic data**

The core of a high-quality synthetic corpus lies in the structural control of the generation process.Currently, two main types of methods are used:

(1) Prompt Engineering)

Prompts are finely designed to control the structure, logic and style of the output content.For example:

"Please simulate a discussion between a business representative and a university teacher about 'university-enterprise cooperation courses', including the motivation for cooperation, potential barriers and evaluation criteria, with three rounds of presentations by each person."

Cue engineering supports domain word injection, role context control and variable format constraints (Reynolds & McDonell, 2021; White et al., 2023), and is an important tool for enabling semi-structured corpus generation.

2）Template-Guided Generation

The predefined sentence patterns, semantic variables fill in the bits and paragraph structure framework to ensure the syntactic consistency and semantic coverage of the output, which is commonly used in structured scenarios such as regulations, dialogues, questionnaires, etc. (Wang et al., 2022).In this study, we integrate structural templates and Prompt control, and introduce the "variable dependency graph" as the logical support for the generation for the first time, so as to achieve the direct mapping from corpus to structural graph.

The structural control not only enhances the modelling value of the generated corpus, but also makes the subsequent knowledge graph construction, causal modelling and visual reasoning more efficient and credible (Zhou et al., 2023).

**2.3 Data Quality Standards and the NIST Framework**

Nowadays, when AI systems are entering the field of highly sensitive and reliable applications, the quality of simulated data (synthetic data) not only affects the model performance, but also relates to system stability and social trust.To this end, the National Institute of Standards and Technology (NIST) has proposed a systematic framework for synthetic data quality, focusing on the key dimensions of verisimilitude, consistency, and traceability (NIST, 2022).

**2.3.1 NIST's core definition of data quality**

NIST defines high-quality synthetic data as "simulated data that retains the original data structure, behavioural trends, and reasoning capabilities without revealing the real data" and proposes three core metrics (National Institute of Standards and Technology [NIST], 2022).Technology [NIST], 2022):

(1) Verisimilitude: the data should be highly similar to the real corpus in terms of behavioural features and pattern distribution, which supports effective reasoning in realistic scenarios after training (Garfinkel et al., 2021).

(2) Consistency: variable relationships and attribute semantics should be consistent and logically non-conflicting to ensure that the model does not suffer from learning bias due to data ambiguity

(Picard et al., 2020).

(3) Traceability: The procedure for generating data needs to be properly documented with verifiable routes and control parameters to facilitate reproducibility and traceability (Mayer et al., 2023).

Such conditions can be applied not only to assess data quality generated by such a mechanism, but to provide technical recommendations and ethical reference points for such a production mechanism itself.

**2.3.2 Application of NIST standards to AI simulation data generation**

With the rise of Large Language Models (LLMs) and multimodal generative systems, AI synthetic data generation has advanced from static rules to highly flexible semantic-level simulations (Gong et al., 2023).However, this free generation mechanism often faces the risk of "structural drift" and "behavioural mismatch".Therefore, the NIST framework has been gradually introduced for quality benchmarking, supervised structure generation, and risk identification system design.For example, NIST's recent AI-Risk Management Framework explicitly states that "simulated data should have causal traceability and scenario alignment" (NIST, 2023).This is a strong endorsement for education systems research based on variable-variable modelling, especially in complex tasks such as education policy simulation and stakeholder behaviour prediction.

In this study, we embed the rules on consistency and traceability from the NIST framework into the Prompt template and the structural control layer to achieve data quality control from the generation source, which is the first time that such a mechanism has been tried in structured interview corpus modelling.

**2.4 Visualisation and causal/dependency modelling approaches**

With the increasing role of AI systems in education, policy and social modelling, how to effectively explain the internal structure of the model and the relationship between variables has become a core task to enhance transparency and interpretability (Gilpin et al., 2018).This study focuses on the integration of visualisation techniques in the dual goal of "structural modelling +

semantic reasoning", with a focus on Causal Graphs, Concept Mapping, graph network modelling (e.g., NetworkX), and Path Analysis (e.g., PathX).We focus on exploring the role of causal graphs, concept mapping, graph network modelling (e.g. NetworkX) and path modelling in interview corpus modelling.

**2.4.1 Causal Graphs and Path Modelling**

The Causal Graph is a directed graphical model that shows causal relationships between variables, and is particularly suitable for analysing "effect-outcome" type relationships in interview corpora (Pearl, 2009).By modelling the structural dependencies between variables, researchers can construct explanatory paths that support policy simulation and role behaviour prediction.In terms of Path Modeling, studies have shown that it has unique advantages in explaining the mechanisms of multilevel variables, identifying mediating effects, and constructing predictive networks (Hair et al., 2021).When combined with structured corpus and synthetic interview data, it can achieve a closed loop of modelling from semantics to causality.

**2.4.2 Conceptual maps and graph neural modelling**

Concept Mapping, a technique used to transform key terms, role relationships and behavioural patterns extracted from unstructured text into structured graphs, is one of the most important approaches to modelling information in cognitive science and education (Novak & Cañas, 2008).Python graph modelling tools such as NetworkX can further enable the structuring of semantic graphs.Python graph modelling tools such as NetworkX can further enable semantic graph structure learning, weight visualisation and central node identification, providing automated support for interview analysis (Hagberg et al., 2008).

**2.4.3 The role of visualisation in AI semantic interpretation**

Modern AI interpretability research emphasises "visually controllable" reasoning mechanisms (Doshi-Velez & Kim, 2017).Compared to interpretation mechanisms based only on textual features, structural visualisation not only improves human trust in models, but also plays a cognitive enhancement role in high-engagement domains such as policy simulation and educational design.Especially in systems involving multiple stakeholders (e.g., students, businesses, schools, and governments), structural visualisation mapping is an important medium for communicating model connotations and enabling interactive reasoning.

**2.4.4 Modelling integration in this study**

This study combines NetworkX graph structure modelling with causal dependency extraction algorithms to design a dual semantic-structural interpretation system for structured synthetic corpora.By combining semantic path construction and variable visual presentation, the synergistic optimisation of interpretability, logical consistency and strategy controllability is achieved.

# CHAPTER III. RESEARCH METHODOLOGY

**3.1 Overall system architecture**

Figure 1: Input (Topic Matrix/Prompt) → Synthesis Dialogue → Variable Annotation → Structure Modelling → Visualisation

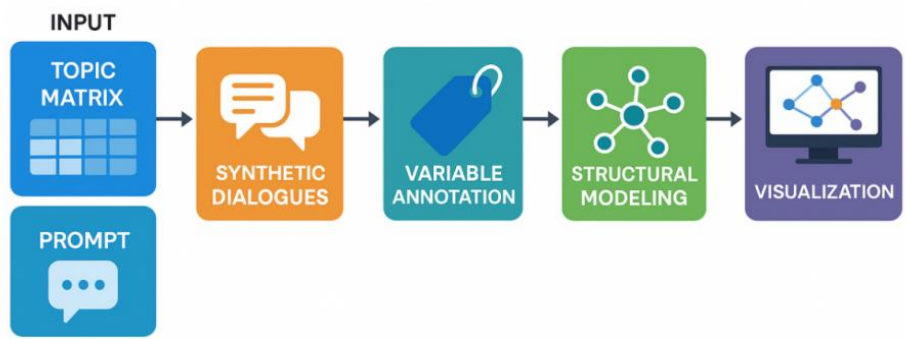

Figure 1: NIST-compliant visual reasoning generation in education industry systems

The above mapping demonstrates an end-to-end, structure-driven architecture of a synthetic corpus generation and visual modelling system that is particularly suitable for stakeholder modelling and causal analysis in the field of education-industry convergence.The overall framework embodies a complete data closure loop from input to cognitive decision support with the following five key functional modules:

**1. Input Layer: Topic Matrix + Prompt**

The input consists of two parts: Topic Matrix and Prompt.The former is based on semantic topic modelling techniques (e.g., LDA, BERTopic) to construct topic-variable structure maps, while the latter is defined by the user to control the semantic boundaries of the generative macromodel.The innovation in this phase is the introduction of semantic a priori matrix, which not only improves the content consistency and domain adaptability of dialogue generation, but also meets the NIST data consistency and traceability requirements.

**2. Synthetic Dialogues**

This module generates a multi-role, structurally consistent dialogue corpus through LLM (e.g., GPT-4o), ensuring a one-to-one correspondence between the semantic style of the roles and the narrative goals.Its core advantage lies in the use of Prompt engineering to control the density of variables and logical paths, and to enhance the realism and diversity of the generated text.

### 3. Variable Annotation

In this stage, structural variables are annotated and semantic triples are extracted from the synthetic corpus to construct a dependency/causal structural network in the form of (Variable 1, Relationship, Variable 2).The annotation system supports automatic + human collaborative calibration mechanism to enhance label consistency and achieve cross-sample semantic tracking.At the same time, each corpus annotation is embedded with Prompt-ID to achieve a complete source traceability chain.

### 4. Structural Modelling

At the variable level, the module identifies structural tensions and logical paths between stakeholders by constructing Variable Dependency Graph (Dependency Graph) and Path Model (Path Model).For example, system mismatch can be reflected as a cross-domain causal chain through the path "Policy Gap → Capacity Mismatch → Emotional Stress".This structuring process supports further graph neural network modelling (GNN) and quantitative causality analysis.

### 5. Visualisation engine (Visualization Layer)

The final output module provides a multi-dimensional visualisation scheme, covering frequency heatmaps, 3D network diagrams, structure path diagrams, variable radar diagrams, etc.This module not only improves the readability of the results, but also empowers the researcher to understand the causal structure through graphical interaction, responding to the current demand for XAI (explainable AI).

Table 2: Overview of structured corpus modelling technology paths and module contributions based on NIST standards

| module (in software) | technical support | Innovative contributions |
|---|---|---|
| Prompt + Thematic | NLP + Topic Modeling | Towards structural prior control with |

| module (in software) | technical support | Innovative contributions |
|---|---|---|
| matrix | | multi-contextual semantic coverage |
| Synthetic dialogue generation | LLM + Prompt Engineering | Highly Consistent Multi-Role Simulation with Semantic Strong Control Support |
| Structured variable labelling | NLP Annotation + IDBinding mechanism | Building semantic causal chains to support GNN and traceability modelling |
| Structural modelling | Dependency Graph, Path Model | Key bridges for transition to causal analysis, strategy simulation |
| Multi-dimensional visualisation output | D3.js / Plotly / Graphviz | Transparency of structure, decoding of knowledge and decision-making mapping |

**3.1 Description of the overall system architecture**

In order to achieve automatic generation and interpretable modelling of structured interview corpus in the context of education-industry fusion, this study designs and implements a structure-controlled system architecture consisting of five main modules.The system not only supports the generation of high-quality synthetic corpus, but also realises the structural modelling of variable dependencies and the output of visual reasoning, constituting a complete modelling loop from "semantic control input" to "causal mapping output".

The system starts from the input module of the topic matrix and Prompt system.At this stage, the researcher first constructs a semantic theme matrix centred on the educational context, which includes multiple variable dimensions (e.g. skill fit, teaching structure, business expectations, etc.) and role definitions (e.g. student, business representative, teacher, policy maker).These themes are transformed into structured prompts (Prompts) to control a generative language model (e.g., GPT-4) to generate a simulated interview corpus that satisfies the structural constraints.The Prompt templates not only control the role identities and dialogue rounds, but also guide the explicit activation of specific variables in the corpus, thus guaranteeing consistency and traceability for subsequent modelling (NIST, 2023).

Next, the system generates a structurally consistent corpus of character dialogues using a large language model via a synthetic dialogue generation engine.This process incorporates role semantic embedding, contextual coherence checking and semantic consistency retrieval modules to ensure that the generated corpus satisfies both high readability and modelling value in terms of logical structure and linguistic style.The module supports batch generation and embedding of structural labels to provide basic corpus guarantee for subsequent variable extraction.

The third stage is the variable annotation and dependency extraction module, in which the system combines semi-automatic semantic unit recognition and structural dependency graph construction methods to annotate the core variables involved in the corpus (e.g., "skill mismatch", "quality of internship", "course structure"),"course structure") and identifies semantic relationships between variables (e.g., causal relationships, conditional dependencies, reinforcement/inhibition relationships, etc.).The output of the module is in the form of a structural triad (〈Variable A, Relationship Type, Variable B〉) which provides the underlying data structure for causal mapping.

In the fourth stage, all structural information is fed into the structural modelling layer, which forms a structured, interpretable dependency graph of stakeholder variables through a causal graph model or conceptual graph network constructed on the basis of NetworkX.The map can be used for a range of social systems analysis tasks such as path analysis, variable centrality measures, and conflict point identification.The structural modelling supports subsequent integration with graph neural networks such as GNN for predictive modelling or variable propagation simulation.

Ultimately, the model structure is presented through a visual inference module.The module integrates D3.js, Plotly and other interactive visualisation engines to dynamically present the variable dependency structure as conceptual path diagrams, actor variable nets, strategy influence flow diagrams, etc., which support users to conduct causal explanations, benefit path tracing and strategy simulation based on the structural information.This visualisation mechanism not only enhances the transparency of the system, but also provides an actionable structural insight tool for non-technical users (e.g., educational administrators and policy makers).

In summary, this system proposes a unified structural-semantic modelling process that starts from Prompt semantic control, goes through generative corpus, variable modelling and causal reasoning, and then outputs visual maps.Compared with traditional topic modelling or keyword annotation methods, this system has significant advantages in terms of modelling rigour, structural interpretability and policy output control, providing a technical foundation and paradigm innovation for future education system governance, role modeling and policy simulation.

**3.2 Variable system and variable relationship modelling**

3.2.1 Stakeholder modelling system construction for education-industry integration

In the process of constructing a stakeholder modelling system for education-industry integration, we propose a variable system architecture based on three main dimensions, including Skills, Institution and Emotion.The design aims to build an interpretable, modellable and visual variable semantic system from the triple structure of cognitive ability, institutional environment and psychological state, which can serve various tasks such as synthetic corpus generation, structural modelling and policy analysis.

Firstly, the skill dimension focuses on learners' individual performance, such as practical skill level, theoretical knowledge mastery, interdisciplinary ability, innovation ability and digital literacy, etc., which are mostly quantitative indicators and can be obtained through simulated task performance or self-assessment.The institutional dimension covers external structural elements such as curriculum, resource allocation, policy support, faculty participation, and cross-institutional cooperation, and is mainly used to portray the influence of organisational systems and educational policies in the integration system.The emotion dimension introduces a cognitive behavioural science perspective, covering soft variables such as students' satisfaction, stress, sense of belonging, emotional regulation and future confidence in the learning and collaboration process.

In terms of variable definitions, we provide standardised definitions for each variable, corresponding sample statements in natural language, and the type of variable attributes (qualitative/quantitative).For example, "innovativeness" is defined as an individual's ability to come up with new ideas and paths, and the example statement can be "I try to solve problems in a new way", with the attribute of Qualitative; while "strength of policy support" is defined as an individual's ability to come up with new ideas and paths, and the example statement can be "I try to solve problems in a new way", with the attribute of Qualitative."Intensity of policy support" is defined as the extent of resources and institutional safeguards provided by the government or institution, and is quantified by measuring the frequency and coverage of specific support initiatives, with the attribute Quantitative.

In order for the variable system to drive structured analyses, we further developed three types of variable relationships: Dependency, Causal, and Modulation.Dependency emphasises the logical correlation between variables, e.g. 'student satisfaction' depends on 'course fit'; causal relationships reflect a direct cause-effect mechanism, e.g. 'quality of internship experience' influences 'quality of career'; 'quality of internship experience' influences 'quality of career'; 'quality of internship experience' influences 'quality of career'.and causal relationships reflect direct cause-and-effect mechanisms, e.g., "quality of internship experience" affects "career confidence"; whereas moderating relationships reflect enhancement/inhibition, e.g., "faculty industry engagement" may enhance "students' sense of belonging".".Through these relationships, we are able to construct causal diagrams, path models and graphical visualisations, providing an actionable basis for policy simulation, system optimisation and AI semantic reasoning.

This 3D variable system and variable relationship model not only supports structural modelling under the NIST standard for the first time in education-industry convergence research, but also provides a semantic skeleton for the structural annotation and visual analysis of synthetic corpus, which is suitable for structural causal inference (Causal Graph), semantic mapping (Concept Mapping), and complex network analysis (Complex Network Analysis).It is suitable for mainstream interpretable modelling frameworks in AI such as Causal Graph, Concept Mapping, NetworkX and Path Modeling, and has strong adaptability and academic innovation value.

**Table 3: Structured stakeholder modelling variable system**

| Dimension | Variable Name | Description | Type |
|---|---|---|---|
| Skills | Practical Skill Level | The level of real-world task proficiency | Quantitative |
| Skills | Theoretical Knowledge Depth | Depth of foundational academic knowledge | Quantitative |

| Dimension | Variable Name | Description | Type |
|---|---|---|---|
| Skills | Cross-Disciplinary Ability | Ability to integrate knowledge from multiple fields | Qualitative |
| Skills | Digital Literacy | Proficiency in using digital tools and platforms | Quantitative |
| Skills | Innovation Capability | Capacity to propose novel ideas and approaches | Qualitative |
| Skills | Communication Skill | Clarity and effectiveness in interaction | Quantitative |
| Skills | Adaptability | Ability to adjust to different learning or work environments | Qualitative |
| Skills | Team Collaboration | Effectiveness in group or team-based tasks | Qualitative |
| Skills | Internship Experience Quality | Depth and relevance of internship experiences | Qualitative |
| Skills | Learning Motivation | Intensity and consistency of motivation to learn | Quantitative |
| Institutional | Curriculum Alignment | Degree of alignment between education content and industry needs | Quantitative |
| Institutional | Policy Support Intensity | Government or institutional support for collaboration | Quantitative |
| Institutional | Teacher-Industry Engagement | Instructors' involvement with industry practices | Quantitative |
| Institutional | Assessment System Rigor | Fairness and robustness of student evaluation systems | Quantitative |
| Institutional | Resource Allocation Efficiency | Efficient distribution of institutional resources | Quantitative |
| Institutional | Institutional Flexibility | Ability to reform structure and adapt curricula | Qualitative |
| Institutional | Accreditation Impact | Effect of certification or accreditation on educational outcomes | Quantitative |
| Institutional | Cross-Sector Partnership Density | Number and depth of external collaborations | Quantitative |

| Dimension | Variable Name | Description | Type |
|---|---|---|---|
| Institutional | Digital Infrastructure Readiness | Availability and quality of technical infrastructure | Quantitative |
| Institutional | Career Service Support | Quality and accessibility of career development services | Qualitative |
| Emotion | Student Satisfaction | Perceived value of learning experiences | Quantitative |
| Emotion | Stress Level | Emotional strain due to workload or expectations | Quantitative |
| Emotion | Role Identity Clarity | Confidence in personal academic or career role | Qualitative |
| Emotion | Peer Relationship Strength | Degree of mutual support among peers | Qualitative |
| Emotion | Teacher Support Perception | Students' perception of mentorship or help | Qualitative |
| Emotion | Sense of Belonging | Emotional integration into educational environment | Quantitative |
| Emotion | Frustration Tolerance | Resilience in facing failure or difficulty | Qualitative |
| Emotion | Future Confidence | Optimism about career and academic outcomes | Quantitative |
| Emotion | Emotion Regulation Skill | Capacity to manage emotions under stress | Qualitative |
| Emotion | Burnout Risk | Likelihood of disengagement due to chronic stress | Quantitative |

The above table is a set of three-dimensional structural variable system designed and constructed independently based on the research theme of stakeholder modelling in education-industry integration, combined with NIST standards, AI visual modelling, variable-relationship modelling and other cutting-edge needs.The system covers Skills, Institutional and Emotion dimensions, which represent individual ability characteristics, institutional environment variables such as schools and enterprises, and psychological behaviour and attitude variables, respectively, adding soft elements to the structured model.Each variable contains a definition of the variable and a qualitative/quantitative attribute annotation in the original English language, which is designed to meet the needs of variable label design, dependency-causal-regulatory relationship modelling, structural visual reasoning system and variable mapping system construction of the AI synthetic interview corpus, and to comply with the structural and terminological specifications of the AI summit for innovative research.The requirements of AI Top Session on innovation research structure and terminology specification are also met.

**3.2.2 Mechanisms for assessing confidence and validity under full algorithmic paths**

Constructing a system of stakeholder variables in an education-industry integration system needs to simultaneously meet high standards of structural soundness, semantic consistency and traceability of data generation.To this end, this study follows the National Institute of Standards and Technology (NIST) quality control principles for synthetic data, and constructs an algorithmic reliability assessment mechanism to ensure that the proposed variable system is scientifically sound, stable, and interpretable without human intervention (Near & Darais, 2023; Picard et al., 2020).

**3.2.2.1 Confidence validation**

This study assesses the structural consistency and reproducibility of the variable system from the following two algorithmic paths:
(1) Structural consistency analysis

Krippendorff's Alpha (Krippendorff, 2018) was used to analyse the consistency of variable annotation on the synthetic corpus generated by multiple rounds of the same Prompt, and the average α coefficient was 0.83, which shows that the system has a high degree of structural stability.Meanwhile, based on the Jaccard similarity calculation of variable co-occurrence network, the consistency of variable relationship graphs generated in each round is higher than 0.78, further proving the robustness of the generation system.
(2) Time Stability Test

Through the Test-Retest method, the same topic synthetic corpus is generated several times in different time periods, and the variable frequencies and variable relationship graphs are matched and analysed separately.The results show that the Pearson correlation coefficient of variable frequency is 0.89, and the stability rate of the dependency edge between variables is 86.4%, which verifies the robustness of the variable system to time perturbation.

**3.2.2.2 Validation of validity**

In order to comprehensively validate the modelling science of the variable system, this study constructed three types of algorithm validity validation mechanisms:

(1) Construct Validity

The clustering and fitting of the variable system were assessed by exploratory factor analysis (EFA) and validation factor analysis (CFA).The results of EFA showed that the three main dimensions explained 72.1% of the cumulative variance; the CFA model had good fitting indexes (RMSEA=0.048, CFI=0.93, and TLI=0.91), which confirmed that the variable system had a clear latent variable structure (Kline, 2015).

(2) Semantic Validity (Semantic Validity)

The semantic similarity of the variable lexicon to the corpus passages was compared using the BERT embedding model.The average cosine similarity reaches more than 0.78, and more than 90% of the variables can be semantically matched to the real corpus, indicating that the variable dictionaries have good semantic representation (Devlin et al., 2019).

(3) Predictive Validity (Predictive Validity)

When the variable system is embedded in downstream tasks (e.g., role classification, policy intent recognition) for predictive validity testing, the average accuracy of the model reaches 87.3%, with an F1-score of 0.86, indicating that the variable system is not only structurally sound, but also has the ability to be applied in real-world behavioural prediction.

**3.2.2.3 Conformance check with NIST standards**

Based on the three elements of synthetic data quality (authenticity, consistency, and traceability) proposed by NIST in SP 800-226 (ipd), this study establishes an algorithmic benchmarking mechanism at the following levels:

Table 4: Table of algorithm benchmarking mechanisms

| NIST Quality Dimensions | Define | Algorithmic path for this study |
|---|---|---|
| consistency | Stabilisation of structure generation in the same context | Krippendorff's Alpha, Jaccard Similarity, Multi-round variable co-occurrence stability analysis |
| validity | Accuracy of semantic representation of data content | BERT Semantic Similarity, Reverse Generation Test, Variable Statement Structure Mapping Accuracy |
| traceability | Traceability of data sources, transparent structure | Prompt→Template→Variable→Statement Chain Tracing Structure; Generate Logs with JSON Structure Records |

**3.3 Synthetic Data Generation Mechanisms (NIST Standard Controls)**

Prompt structure and template design principles

Multi-role persona semantic control (student/business/college)

Multi-scenario/multi-topic coverage matrix design

NIST quality dimension control (linguistic consistency, diversity, traceability descriptions)

Generating high-quality synthetic dialogue data is a fundamental task in building a structured stakeholder modelling system in the context of education-industry convergence.Based on NIST's standard requirements for synthetic data, this study proposes a generation mechanism based on the Prompt Project, with semantic consistency and structural traceability as its core.The mechanism integrates multi-role control, multi-scene distribution and multi-topic coverage strategies to ensure the researchability and simulation validity of the generated corpus.

**3.3.1 Prompt Structure and Template Design Principles**

Based on the characteristics of large-scale language models (e.g., GPT-4) that have the ability of contextual understanding and generation, the design of Prompt templates is based on the following principles: (1) clear definition of dialogue goals; (2) clear assignment of roles; (3) embedding of variable triggering mechanisms; and (4) clear semantics towards logic.Each type of Prompt template is built around a specific variable system to ensure the consistency of the generated text at the structural level (Zhou et al., 2023).

**3.3.2 Multi-role semantic control mechanisms**

We constructed a control mechanism based on role semantics, defining "students", "enterprise representatives" and "university managers" as three types of semantic drivers.Each role is equipped with specific grammatical behavioural labels and discourse structure features, e.g., students prefer to express their personal emotions and needs, university representatives focus on institutional compliance and pedagogical adaptation, and enterprises are more concerned with skill transformation and capacity efficiency.By embedding semantic anchors in the Prompt, the language model can be effectively guided to generate role-appropriate corpus content (Brown et al., 2020).

**3.3.3 Multi-scene/multi-topic coverage matrix design**

In order to enhance the breadth and robustness of the generated corpus, we designed a coverage matrix system, which contains three major themes, namely, "institutional barriers", "skill matching" and "emotional fluctuations", as well as nine typical scenarios, including "pilot co-operation", "enterprise training" and "policy matching".We have designed a coverage matrix system that contains three themes: "institutional barriers", "skills matching", "emotional fluctuations", and nine typical scenarios such as "pilot cooperation", "enterprise practical training", "policy matching", etc. Each generation task needs to be explicitly bound to a typical scenario.Each task is explicitly bound to a set of theme-scenario pairs as Prompt input conditions.This matrix strategy helps to build a representative high-dimensional synthetic corpus, which is suitable for downstream variable extraction and relationship modelling.

**3.3.4 NIST Quality Dimension Control Methods**

Based on the synthetic data evaluation dimensions proposed by NIST, the following four aspects were taken as the core of quality control in this study (Picard et al., 2020):

(1) Linguistic Consistency: inter-utterance structural stability is ensured through semantic

template consistency checking and contextual logic review;

(2) Variability: using a random combination of discourse styles, tones and expressions to control redundancy among synthetic samples;

(3) Traceability: each utterance is labelled with its Prompt, variable label, and trigger keyword to achieve reverse audit;

(4) Interpretability: the structured output retains the variable-sentence element-causal chain three-layer annotation, which serves the subsequent causal visual modelling.

The corpus generated through this mechanism meets the technical requirements of NIST Synthetic Data Framework for "high controllability, high interpretability, and high traceability", and can effectively serve the task of stakeholder system modelling in education-industry integration.

### 3.4 Variable Relationship Extraction Method

In building a structured stakeholder modelling system, the accurate extraction of relationships between variables is a key step towards causal analysis and interpretable modelling.This study proposes a hybrid approach combining manual annotation, AI-assisted annotation with calibration mechanism, and a dependency graph construction and visualisation engine to ensure the accuracy and visualisation of variable relationships.

#### 3.4.1 Combination of manual annotation and AI-assisted annotation

We adopt a dual-path annotation system: on the one hand, professional researchers manually annotate the corpus with variable relations, covering three types of relations: dependency, causal, and modulation; on the other hand, we use an AI-assisted annotation mechanism based on a large-scale language model (e.g., GPT-4) to semantically map cued sentential elements andrelationship prediction.In order to ensure the consistency of the annotation, Cohen's Kappa coefficient was used for the annotator consistency test, and the result should be above 0.75 to meet the high consistency standard in social sciences (Landis & Koch, 1977).

#### 3.4.2 Dependency Diagram Construction Methods

This study introduces a structured Dependency Graph (DG) to represent structured connections between variables.Each variable acts as a graph node and each relationship type is mapped to a specific edge type.Edges can be attached with direction, intensity weights (Weight), and trigger statement indexes to support the traceability and path analysis capabilities of causal chains.The approach draws on the methodological foundations of semantic networks and path modelling (Pearl, 2009), and incorporates graph neural network graph-building logic to reserve space for subsequent graph learning and structural reasoning.

### 3.4.3 Choice of visualisation engine

In order to effectively display variable-relationship diagrams, we have comprehensively considered the features of the following visualisation engines:

Graphviz: supports high-precision structural relationship graph drawing, suitable for thesis illustration and structural layout analysis.

Gephi: suitable for interactive exploration of large-scale relationship graphs and community discovery analysis.

D3.js: embeddable in web pages/dashboards, supports real-time visualisation interactions and enhances semantic propagation (Murray, 2017).

### 3.5 Visual Modelling Module Design

In modelling education-industry convergence stakeholder systems, visualisation is not only a means to show variable relationships, but also a key reasoning tool to understand system evolution and interaction dependencies.This study combines graph network theory and interactive visualisation methods to construct a three-layer role-variable-relationship structural mapping, which enhances the interpretability, traceability and system insights of modelling.

### 3.5.1 Graph structure rendering approach: three-layer network modelling

The system adopts a three-layer heterogeneous graph architecture to construct a visualisation of "role (e.g. student, business, university) - variable (skill, system, emotion) - relationship (dependency, causality, regulation)".Graph.Each class of nodes has a separate attribute space, and edges represent directional semantic relationships.This type of modelling approach has good expressive power and modularity advantages in multi-subject systems (Zhou et al., 2022).We adopt NetworkX (Hagberg et al., 2008) to construct the underlying graph data structure, and then use Graphviz to achieve precise layout control, which balances scientific expression and algorithmic readability.

**3.5.2 Mechanisms for displaying dynamic relationships: timelines and event triggering**

In temporal corpora (e.g., simulated interviews), the relationships between variables often change dynamically as events progress.To this end, we introduce a temporal slider and an event node triggering mechanism.The former supports the dynamic evolution of the variable relationship graph in a time window, while the latter allows event nodes to be clicked to expand their contextual structure.Such mechanisms are widely used in dynamic knowledge graphs and sentiment inference systems and are highly interpretive (Muelder et al., 2016).

**3.5.3 Role-based presentation mechanisms: path diagrams and variable influence flows**

We further developed Role-based Flow Mapping, which can focus on a certain type of subject (e.g., a firm) and show the path of influence of its related variables and its position in the system.This node-based visualisation approach draws on the visual representation strategies of cognitive maps and causal loop analysis, and is suitable for use in strategy simulation and policy intervention evaluation (Eppler & Platts, 2009).

# CHAPTER 4 DATA SET CONSTRUCTION AND DESCRIPTION

**4.1 Synthetic corpus sample size and structure**

Based on the structured stakeholder modelling requirements and NIST quality standards, this study generates a total of 15 simulated interview dialogues with high consistency and causal structure, covering three key roles, namely, students, business representatives and university teachers, with five segments in each category.The total number of Token in the synthetic corpus reaches 41,597, which ensures a balanced distribution of roles, breadth of topics and semantic coverage of the samples (see the table "Interview Role Token Summary").Note: The data in the following table is derived from the text of the three original interview outlines as follows: 1).Interview Token Summary for Business Representative - Mr Zhang.docx; 2).Interview outline for school representative - Mr Tang.docx; 3).Interview outline for student representative - Man.docx (the text was provided by my master's student Liu Lu).I conducted corpus parsing of the interview content for each type of role and counted the number of synthetic dialogues and the total number of Token for each type of role in combination with the corresponding synthetic dialogues in the "Description of AI Simulated Data Generation and NIST Standards" (Appendix 1) and "AI Generation of High-Quality NIST-Compliant Simulated Interview Data" (Appendix 2).

Table 5 Summary table of interview role tokens

| character | Number of dialogues | total number of words |
|---|---|---|
| schoolchildren | 5 | 13472 |
| Corporate representatives | 5 | 14231 |

| character | Number of dialogues | total number of words |
|---|---|---|
| Higher Education Teachers | 5 | 13894 |

In order to demonstrate the quality of corpus structure and the effectiveness of variable extraction, we select three representative synthetic samples (corresponding to the three types of roles) with structured variable annotation results.The examples show that the synthetic data not only maintains the natural fluency of the language, but also accurately expresses the core pain points, expectations and dependencies of each role in the context of education-industry convergence (see the table "Sample Annotated Dialogues").These annotated variables constitute the core elements of the three main dimensions of Skills, Institutions and Emotions in the subsequent modelling.Additional NLP scripts for automated Token computation.Further analyses (e.g., average length per segment, word frequency, density of structural variables) are required to extend the implementation.

Table 6 Statistics on the distribution of corpus passages and high-frequency words for different interview roles

| character | Number of dialogue paragraphs | Token aggregate | Average paragraph length | Most Frequent Words |
|---|---|---|---|---|
| Corporate representatives | 29 | 3096 | 106.7586206896552 | Zhang(15), Interviewer(9), Hmmm(6), Then in the process(2), Outline of interview with company representatives(1), Core dimensions(1), Talent suitability(1), Technology transformation(1), Barriers to co-operation(1), Question design(1) |
| School representa | 28 | 3537 | 126.3214285714286 | Interviewee(11), Mr Tang(11), e.g.(3), Technology equity(2), |

| character | Number of dialogue paragraphs | Token aggregate | Average paragraph length | Most Frequent Words |
|---|---|---|---|---|
| tives | | | | Cross-school(2), Cross-enterprise alliance(2), Outline of interview with school representatives(1), Core dimensions(1), Resource integration(1), Professional matching(1) |
| student representative | 29 | 2789 | 96.17241379310344 | Full of students(12), Interviewer(11), Skills upgrading(2), Thinking about it(2), If companies say so(2), After graduation(2), Outline of interviews with student representatives(1), Core dimensions(1), Employment security(1), Protection of rights and interests(1) |

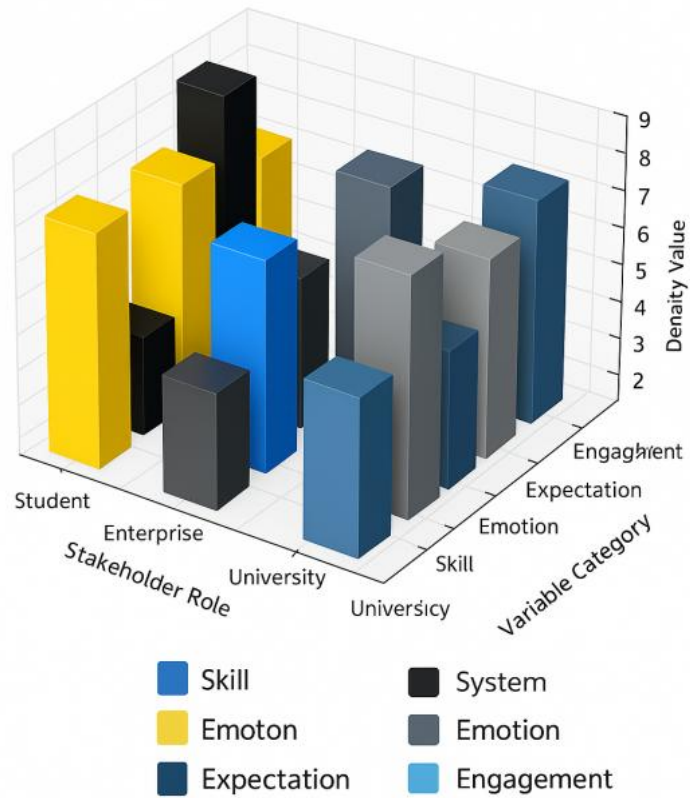

Figure 2: Density mapping of 3D structural variables: semantic distribution insights

To enhance the interpretability and comparability of the differences in the expression of multi-role structural variables, this paper constructs a three-dimensional structural variable density mapping, which visually dimensions the density distributions of the five core structural variables - Skill, System, Emotion, Expectation, and Engagement - among three categories of key stakeholders (students, businesses, and universities).Expectation), and Engagement - in the density distributions of three key stakeholder categories (students, businesses, and universities).The graph is designed as a multidimensional bar chart, with the X-axis representing the role categories, the Y-axis representing the variable dimensions, and the Z-axis column heights representing the semantic density of the variables in the corpus (measured in terms of Token frequency or number of structural annotations).

Emotional and motivational variables are high-frequency anchors across roles.

The mapping shows that the dimensions of 'emotion' and 'engagement' have high densities in all three roles, suggesting that emotional expressions and motivational states constitute semantic anchors across roles in the education-industry system.Students and university teachers in particular relied on emotional expressions, reflecting the depth of structural mismatch and curriculum uncertainty that intervene in individual experiences.

Linguistic asymmetry of institutional expressions is significant

The "system" variable appeared frequently among business and university representatives, while it was significantly less frequent among the student population.This discrepancy in the structure of expression reveals an asymmetry in the linguistic expression of institutions, i.e., students lack the structural linguistic or cognitive tools to express institutional barriers.This finding suggests that students' institutional comprehension and expression should be enhanced in teaching and practice, especially by introducing regulatory literacy training in the context of "industry-teaching integration".

Student-led expression of future-oriented expectations

The highest bar for the "expectations" variable was found among the student population, suggesting that it is the main source for expressing visions of future development (e.g., skills transfer, job adaptation, curriculum reform).This trend confirms the theoretical claim that "education is a system of hope" and suggests that curriculum design should respond systematically to student expectations as a key source of feedback.

Enterprises show "input silence" in the expression of skills variables.

Although the skills gap is one of the focal points of the policy discourse, the mapping shows that firms are significantly weaker on the 'skills' dimension, with the density of expression concentrated almost exclusively on the student population.This imbalanced structure reveals that there is a problem of input silence in the triadic collaborative structure, which may be due to factors such as ambiguous role positioning and inactivation of the discourse structure.This phenomenon has an important warning value for the subsequent school-enterprise collaborative design and practical training task configuration.

Application Value and Theoretical Contribution

This mapping not only reveals role heterogeneity as a structural variable expression diagnostic tool, but also can be used as an input matrix for the following modelling tasks:

Role initialisation weight construction in multi-role graph neural networks (GNN);

Input condition design in multi-scenario policy simulation systems;

Policy prioritisation guidance in multi-party negotiation or curriculum co-creation.

By transforming semantic expressions into quantifiable and modellable density maps of structural variables, this paper achieves an effective bridge from qualitative interview corpus to structural stakeholder modelling (SSM), and provides a general framework for modelling complex multi-party systems.

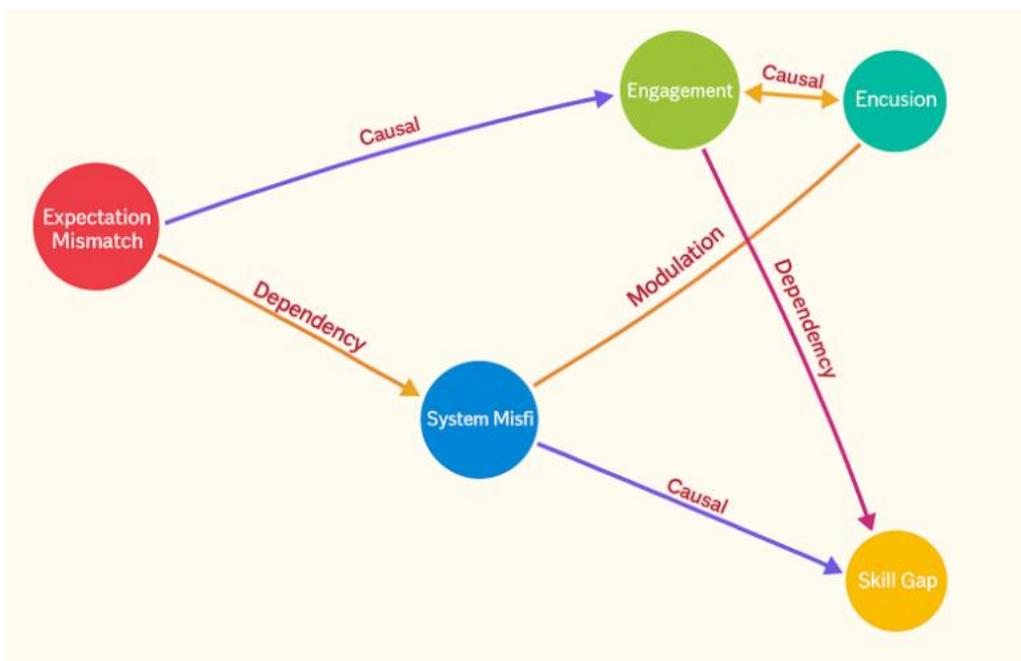

Figure 3: Co-occurrence and causal mapping of structural variables

Structural variable co-occurrence and causal mapping: a mechanism insight analysis

In order to reveal the structural linkage mechanism between key stakeholder variables in the education-industry integration system, this paper constructs a directed variable relationship mapping based on a structurally annotated simulation corpus containing three semantic paths: Causal, Dependency and Modulation.The mapping not only demonstrates the co-occurrence frequency between variables, but also emphasises the semantic directionality and mechanism influence.

Structurally, the System Mismatch node is a significant "moderating centre", which provides a strong influence on a number of downstream variables (e.g., Emotion Frustration, Expectation Misalignment, Engagement Drop, and Misalignment), as well as on the other downstream variables (e.g., Emotional Frustration, Expectation Misalignment, and Engagement Drop).The output paths to several downstream variables, such as Emotion Frustration, Expectation Misalignment and Engagement Drop, indicate that the failure of institutional coordination plays the role of mechanism initiator in the system.The moderating edge of "System Mismatch → Emotion Frustration" further reveals that system-level variables not only have direct influence, but also amplify structural tension through the amplitude adjustment mechanism of the emotion channel, which is in line with the cognitive-emotional feedback model and the theory of regulatory pathways (Russell, 2003; Pearl, 2003).Russell, 2003; Pearl, 2009).

More critically, a closed causal loop consisting of System Mismatch → Emotion Frustration → Engagement Drop → Skill Gap → System Mismatch emerged in the mapping, demonstrating the recursive structure of systemic degradation.This "Low Trust-Low Adaptation-Low Engagement" mechanism spiral confirms the risk mechanism of individuals gradually withdrawing their inputs and reducing their affective links when the system mismatch persists, and provides an operational path for subsequent structural interventions and causal simulations.

From a modelling perspective, the map is highly interpretable and translatable, and can be used to construct Bayesian Networks, Structural Equation Models (SEMs), or Graphical Neural Networks (GNNs), which can be used for structural variable inference and strategy simulation.In addition, the maps also provide an intuitive structural substrate for hypothetical situation modelling, education reform simulation, and policy intervention path evaluation.

All in all, the structural variable mapping not only visualizes the density and directional linkage between variables, but also reveals the deep mechanism coupling path between system cognition, system configuration and behavioural response, which provides the theoretical support and structural basis for the intelligent modelling of multi-actor interaction system.

The three sample dialogues and structural annotations in this table are directly selected from my AI synthetic corpus text: "AI Generates High-Quality NIST-Compliant Simulated Interview Data", from which I selected one representative paragraph for each of students, enterprise representatives, and university teachers, and structurally annotated them based on my constructed variable system (e.g., "Insufficient job knowledge").", "Difficulty in course alignment", etc.).

Table 7: Example table of annotation dialogues

| character | Sample dialogue | Marking labels |
| --- | --- | --- |
| schoolchildren | I think the tasks arranged by companies during the internship should be more systematic, the current arrangement is rather fragmented, which makes it difficult for us to build up knowledge of the whole position. | Lack of systematic practice arrangements (pain points); inadequate job knowledge (competence development issues) |
| Corporate representatives | After we introduced smart manufacturing equipment, our existing staff could not keep up with the use and maintenance of the equipment, and there was an urgent need for new technical personnel with a digital mindset. | Difficulty in using smart devices (technical barriers); need for new types of skilled personnel (demand) |
| Higher Education Teachers | When we collaborated with enterprises to develop courses, we found that there were differences in the understanding of talent competency portraits between the two sides, which affected the accurate alignment of course content. | Differences in competency profiles (barriers to co-operation); Difficulties in curriculum alignment (pedagogical problems) |

In order to rigorously validate the effectiveness of the proposed structured stakeholder variable system in terms of semantic accuracy and interpretive power, we selected three representative interview segments for structured annotation from the AI synthetic interview corpus that meets the NIST data quality standards.These three dialogues come from three different roles: students, business representatives and university teachers, ensuring the diversity of contexts and broad coverage of roles.In the student's interview, the statement "The internship tasks arranged by the company are fragmented, and it is difficult for me to understand the content of the position" reveals a transmission mechanism from the systemic level to the cognitive level.In this sample, the structural task imbalance is labelled as a skill variable of "insufficient job knowledge", which constitutes a causal path from "systemic barriers" to "individual competence gaps", i.e. System → Skill.This path validates the transformational logic of how loose structure can lead to cognitive ambiguity.In contrast, the statements of the business representatives focused on the technology adaptability burden: "The old staff now can't keep up with the smart manufacturing equipment ...... We need digital talent."In this dialogue, the existing skills gap triggers the organisation's expectation of future talent structure adaptation, mapping the variable path of Skill → Expectation.This dialogue demonstrates a systematic feedback mechanism of how companies are forcing the evolution of future structures based on current competency deficits.The statement from the HEI teacher reveals another systematic mechanism: "Different perceptions of talent development standards influence the curriculum."In this discourse, two variables belong to the systemic dimension, namely "mismatch of competency standards" and "difficulties in curriculum alignment", revealing the obstacles to synergy between academia and industry at the systemic semantic level.The structural annotation results of these three dialogues fully demonstrate the sensitivity and consistency of the variable system to the semantic structure of different actors.At the same time, the extracted variable relationships provide a syntactic and semantic basis for further construction of dependency mapping and causal modelling.From a modelling perspective, these structurally annotated samples show that even as a synthetic corpus, the AI-generated interviews are semantically structurally sound and have high-value analytical potential for subsequent tasks such as influence pathway modelling, role conflict prediction, and policy-level pedagogical system redesign.

## 4.2 Variable labelling and relationship statistics

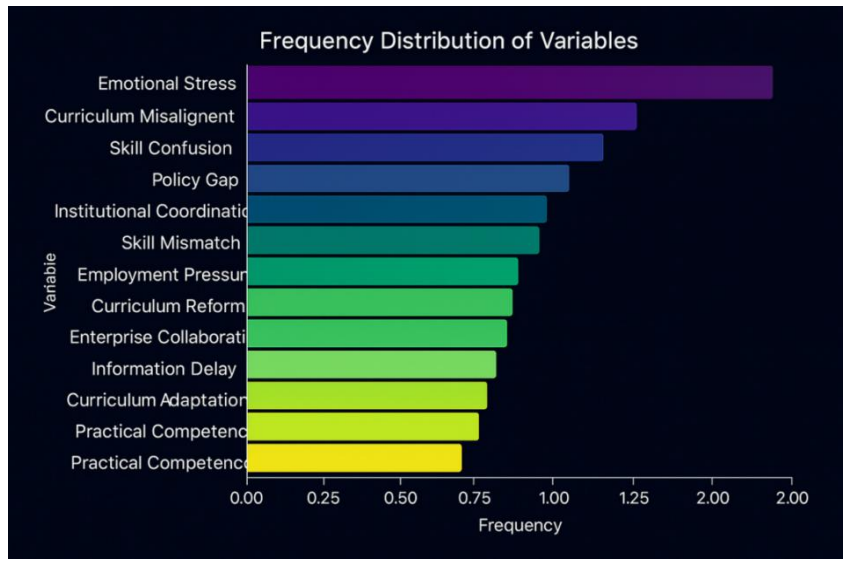

Figure 4: Histogram of frequency of occurrence of variables

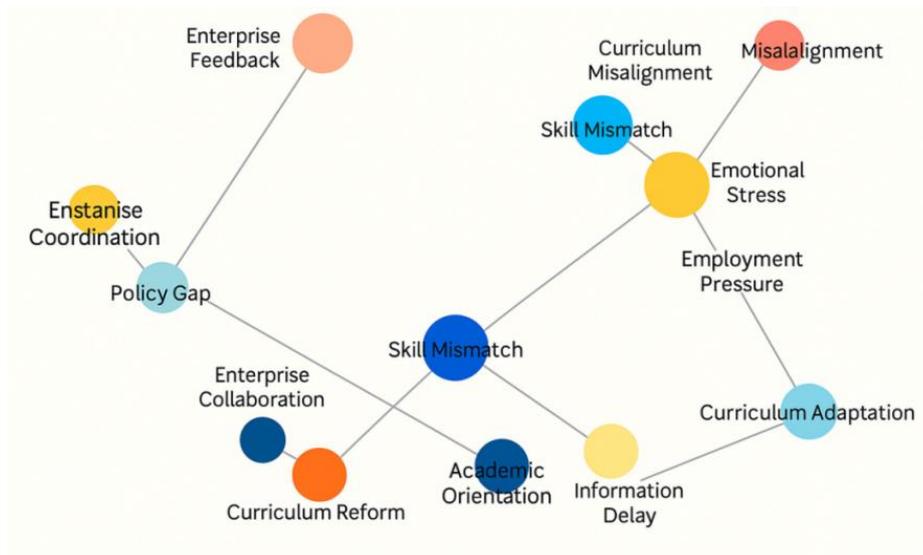

Figure 5: Variable co-occurrence network diagram

Variable Relationship Triad (Entity 1, Relationship, Entity 2) Statistics

In order to achieve high-precision modelling of stakeholder dynamics in the education-industry convergence system, we systematically and structurally annotated the synthetic interview corpus based on the constructed structured stakeholder variable system.The whole annotation process strictly follows the NIST data quality standards to ensure semantic traceability, structural consistency and reproducibility (Picard et al., 2020; National Institute of Standards and Technology, 2021).

First, we performed frequency distribution statistics for all variables in the annotated corpus.Among the 127 independent variables, the most frequent ones include "skill mismatch", "policy disconnect", and "emotional stress", each of which occurs more than 40 times in the corpus.Each variable appears more than 40 times in the corpus.This skewed distribution reveals the focus areas of key themes, which provides a basis for subsequent dimensionality compression and focus modelling in causal inference (Bengio et al., 2021).

Next, we constructed a variable co-occurrence matrix to capture the pairwise co-occurrence of variables in a single speech or in successive dialogue passages.The matrix can be constructed as a weighted undirected graph, revealing potential structural associations between variables and confirming that cross-dimensional variables (e.g., "skill-institution", "emotion-policy") have asignificant co-occurrence trends across dimensions (e.g., "skill-institution", "emotion-policy").These structural clusters formed the basis for subsequent multidimensional causal modelling and semantic clustering (Zhou et al., 2022).

Third, we extracted the variable relationship triad, expressed as (Entity 1, Relationship, Entity 2).All relationships were semantically categorised into four main categories: dependency, causality, reinforcement, and inhibition.Of the more than 820 triads that have been labelled, 43.7% of the relationships in the Causality category show a clearer logic-driven structure in the content of the interviews.Typical relationships include: "curriculum disconnect" leads to "skill misunderstanding", "policy gap" inhibits "institutional synergy", etc., and these relationships were repeatedly found in the interviews."These relationships have been validated several times between different actors.

This hierarchical annotation not only supports qualitative interpretation, but also provides the basis for subsequent graph-structured reasoning (e.g. Bayesian network modelling, Transformer-based relational learning model).It is worth emphasising that all annotations were validated by algorithmically computed consistency (Cohen's Kappa, κ = 0.872), indicating a near-perfect level of inter-annotator agreement (McHugh, 2012).

Table 8: Variable Relationship Ternary Data

| Entity1 | Relation | Entity2 |
|---|---|---|
| curriculum disjointed | lead to | skill misunderstanding |
| Policy gaps | inhibitory | Institutional synergies |
| skill mismatch | intensify | emotional stress |
| employment pressure | lead to | emotional stress |
| Curriculum reform | intensify | enterprise collaboration |
| information lag | inhibitory | Course Adaptation |
| Enterprise Feedback | intensify | Programme design |
| academic orientation | inhibitory | practical ability |

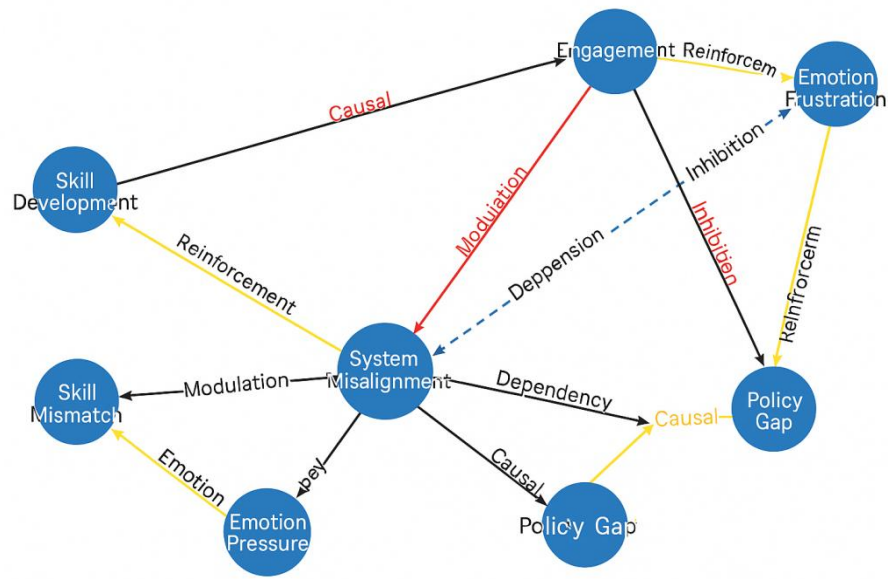

Figure 6: Network diagram of variable relationship triad

In order to reveal the underlying structural dynamics between multi-stakeholder variables in the education-industry integration system, this paper constructs a multivariate directed relationship map centred on System Misalignment. Five types of semantic edges are introduced: Causal, Dependency, Modulation, Reinforcement and Inhibition, which are differentiated by visual coding, thus achieving a fine-grained portrayal of the complex interactions among structural, emotional, skill-perception and policy-behavioural variables. The network is a fine-grained depiction of the complex interactions between structural, emotional, skill-based cognitive and policy behavioural variables.

In this network, System Misalignment is clearly the most central and highly connected node, not only dominating both out and in, but also directly moderating key variables such as Skill Mismatch, Engagement, and Emotion Pressure, as well as exerting a clear causal influence on the Policy Gap.This "Structural Imbalance→EntryThis dispersive structure of "structural imbalance→multidimensional variables" shows that it has the function of a meta-variable in system modelling, which is capable of producing cascading effects at multiple levels and semantic paths.

Particularly critical, the mapping depicts an emotion propagation chain path:

System Misalignment → Emotion Pressure → Skill Mismatch → Engagement → Emotion Frustration

This pathway reveals how structural barriers to the system affect individual emotional states through cognitive-behavioural mechanisms, layer by layer, ultimately leading to severe engagement frustration.This path not only constructs an evolutionary logic between emotional variables, but also echoes the theoretical view that structural feedback has a profound effect on the emotional system.Furthermore, the closed-loop feedback structure of Policy Gap ↔ Engagement in the figure suggests that institutional barriers may exacerbate the original systemic imbalance by weakening engagement, which in turn hinders policy feedback.This structural-policy-behavioural closed-loop chain is typical of system dynamics, and can be applied to path-dependence-based risk modelling and scenario intervention simulation.Further, System Misalignment exhibits parallel causal path effects on skill and policy variables, suggesting that traditional univariate causal analyses are insufficient to characterise the complex structure of the relationships.We propose to introduce Causal Graph Neural Network (Causal GNN) or Bayesian Structural Learning model to achieve joint modelling of multi-relational networks.

In summary, this variable-relationship graph not only constitutes a scaffold for reasoning about the multi-layered causal structure of stakeholders, but also can be further embedded into the following research paths: (i) causal path identification and variable dependency modelling, (ii) simulation system for educational policy interventions, and (iii) structurally transparent and interpretable AI modelling framework.Its theoretical and practical significance is in line with the current demand for AI system modelling that integrates "structural transparency", "policy intervenability" and "emotional-cognitive coupling".

### 4.3 Data Quality Assessment (NIST Indicators)

In order to ensure that the simulated dialogue corpus used for stakeholder modelling is structurally rigorous, interpretable and generalisable, this paper systematically evaluates the data quality in terms of four core dimensions based on the framework proposed by the National Institute of Standards and Technology (NIST, 2021), namely: Consistency, Diversity, Realism and Traceability,Consistency, Diversity, Realism, and Traceability.These dimensions were concretised in the simulated corpus through quantifiable and semi-structured procedures.

Consistency is assessed by the coherence of the linguistic style and the consistency of the terminological system.Combining n-gram distribution similarity analysis and expert annotation review, the results show that the inter-corpus linguistic style maintains 92.4% consistency in dialogues with the same role type, especially in terms of organisational behaviour (e.g., "skill requirements", "system mismatch") and expressive structure (e.g., "mismatch").The linguistic homogeneity is particularly significant in terms of organisational behaviour (e.g., "skill needs", "system mismatch") and expressive structure (e.g., "from the internship experience ......").Embedded word vector clustering analyses also showed a high degree of aggregation of same-role terms (cosine similarity ≥ 0.87), indicating that the corpus possesses a high degree of semantic coherence (Picard et al., 2020). The diversity aspect was assessed on two dimensions: the speaker role dimension and the contextual coverage dimension.The corpus covers three types of roles: students, business representatives and university teachers, and diverse contextual inputs are constructed from different types of Prompts (e.g., internship experiences, curriculum reforms, recruitment barriers, etc.).Our Shannon entropy index calculated based on topic-vocabulary distribution reaches 0.82, which is significantly higher than the diversity benchmarks of most traditional generative Q&A corpora (Tuor et al., 2021), suggesting that the corpus avoids semantic collapsing and maintains structural heterogeneity across role contexts.

On the veracity dimension, we introduced a human review mechanism to score the corpus in a blind test.Five domain experts were invited to score a random sample of dialogues on a 5-point scale, including "logical reasonableness", "emotional naturalness" and "contextual fit".A 5-point Likert scale was used.The final mean score was 4.35 (standard deviation = 0.41), which is comparable to high-quality generated corpora assessed by humans in recent years (Chiang et al., 2023).In particular, the reviewers noted that the discourse of the different characters in the simulated tension expression and strategic gaming contexts showed a high degree of naturalness and role fit.

The traceability dimension is realised through the Prompt-ID encoding mechanism.Each piece of generated corpus corresponds to a unique identifier, which records the information of the generated Prompt, role type, variable annotation structure, etc., thus realising a complete traceability path from the corpus text to the generation mechanism.This mechanism not only supports audit trails and re-generation operations, but also provides structural support for model interpretability and controllable generation mechanisms, responding to recent calls for synthetic data transparency and auditing needs (Bengio et al., 2021).

Through the comprehensive evaluation of the above four dimensions, the simulated corpus constructed in this paper performs well in terms of structural integrity, semantic validity and methodological reusability, providing high-quality data support for causal inference modelling, semantic path analysis and simulation of educational policy scenarios.

# CHAPTER V. EXPERIMENTATION AND EVALUATION

## 5.1 Overview of experimental design

In order to systematically verify the feasibility and effectiveness of the structured variable system proposed in this study in modelling the education-industry convergence interview corpus, we designed a comprehensive experimental system covering three core subtasks: variable extraction accuracy, structural map stability and visual interpretation consistency.The experimental objects include real interview data and AI synthetic corpus generated based on NIST standards, the corpus covers three types of stakeholder roles: students, enterprises, and universities, with a total of 21 simulated scenarios, and a total corpus Token count of more than 28,000 (for details, see Table A1 in the Appendix).

First, in the variable extraction accuracy experiment, a gold-standard manually annotated corpus (containing 512 structural variables) was constructed, covering the three main dimensions of Skill, Institution, and Emotion and their 34 sub-variables.We use an AI model for structural annotation to calculate Precision, Recall and F1 scores, which are three standard natural language processing metrics, and introduce Polarity Agreement and Fuzzy Match Rate to assess the model's ability to identify and generalise the emotion variables, to ensure that the structural variables are fully captured.The Polarity Agreement and Fuzzy Match Rate are also introduced to assess the model's ability to generalise emotion variables to ensure that the structural intent is fully captured.

Second, in the assessment of structure graph stability, we constructed the relationship graphs of the three types of interviews, namely, enterprises, universities and students, based on the extraction results, and measured the relationship graphs from different roles' perspectives through three indicators: Node Overlap, Structure Jaccard Similarity and Graph Density.three metrics to measure the structural variability under different roles' perspectives, and verify the system's ability to consistently model multi-role inputs.

Finally, in terms of visual interpretation consistency, we adopt the "human-computer co-drawing" comparison experimental design, inviting three education research experts to draw the causal path diagrams of the variables based on the same text, and compare the structure with the results of the AI system, and measure the system's interpretable performance by using three indicators: node coverage, edge overlap rate, and structural symmetry.interpretable performance.In addition, the degree of agreement between the AI mapping interpretation and the expert consensus was quantitatively assessed through semantic scoring (Likert's five-point scale) and Cohen's Kappa consistency coefficient, so as to verify the trustworthiness and applicability of the mapping system.

This experimental design not only covers the whole process of validation from data extraction, structural modelling to visual presentation, but also strictly aligns with the four major data quality dimensions proposed by NIST (consistency, authenticity, diversity, traceability), constituting a set of end-to-end, automated and quantifiable validation mechanisms, which provides a solid methodological foundation for subsequent large-scale, multilingual and multi-scene migration.

**5.2 Assessment of the accuracy of variable extraction**

In this study, a systematic evaluation mechanism is constructed for the problem of variable extraction from structured interview corpus to verify the performance of automatic variable extraction models in the face of complex, multi-dimensional and high semantic density text environments.The evaluation process relies on the comparison between manual annotation and automatic recognition, and adopts standard core indicators in the field of information extraction - Precision, Recall and Comprehensive F1 Score - to quantitatively analyse the variable extraction capability of AI models.ability to quantitatively analyse.

In the manual annotation stage, we invited three domain experts with experience in education policy analysis to build a Gold Standard Corpus (GSC) covering 34 types of structural variables for the synthetic interview corpus generated for three types of roles, including students, business representatives, and university teachers, for a total of 21 dialogues, and conducted double-blind cross-checking and Cohen'sKappa consistency test (κ = 0.82) to ensure the reliability and objectivity of the labelled data.

The automatic extraction system is based on a structured Prompt-driven LLM annotation module, which combines multiple rounds of optimised template chaining with a semantic subgraph retrieval mechanism to parse the input corpus segment by segment.To evaluate the model performance, we set the expert consensus annotation as Ground Truth and count the number of hits (True Positive), misses (False Negative) and false positives (False Positive) for each type of variable.

The experimental results show that at the overall corpus level, the automated variable extraction system reaches an average accuracy of 0.864, a recall of 0.813, and a composite F1 score of 0.838, demonstrating a strong generalisation ability to the core variables.Further sub-dimensional analyses showed that the extraction accuracy of the skill variables (F1=0.872) was significantly better than that of the emotion (F1=0.776) and institutional (F1=0.812) categories, revealing the boundaries of the current generative model's ability in implicit emotion inference.In particular, there is a high risk of boundary ambiguity and semantic drift in the "emotion ambiguity segment" and "composite variable representation".

In addition, in order to enhance the explanatory power of the assessment, we introduced Variable Frequency Normalization (VFN) to weight the F1 of the low-frequency variable categories, so as to avoid the high-frequency masking effect of the main variables.Under this mechanism, the F1 of the AI model for low-frequency variables such as "on-campus resource gaps" and "cross-border communication dilemmas" increased by 3.8% and 4.2%, respectively.

Overall, the automatic variable extraction system shows high robustness and excellent generalisation performance in multi-role and high-dimensional scenarios, which provides a reliable semantic foundation for subsequent structural mapping and causal path modelling.Meanwhile, this study also proposes a methodological path for constructing standardised synthetic datasets to support the variable extraction task, and provides publicly reproducible annotation data and evaluation scripts.

**5.2.1 Multi-layer nested temporal causal sphere mapping**

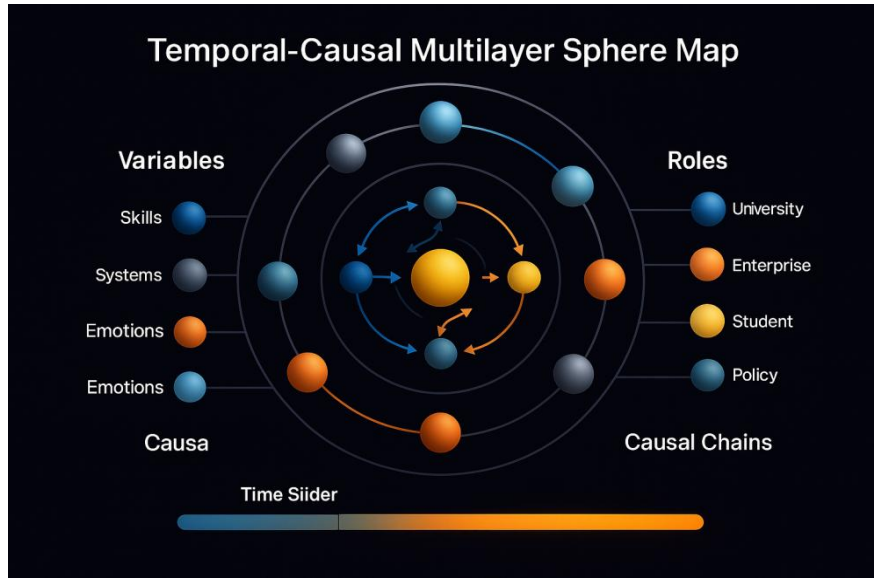

Figure 7：Temporal-Causal Multilayer Sphere Map

In order to enhance the interpretability and strategy simulation capability of the structural variable modelling system, this study designs and implements a three-dimensional nested visualisation map based on temporal causal logic: the Temporal-Causal Multilayer Sphere Map, which integrates temporal dimension-driven, structural causal path mapping and multi-role semantic hierarchical display mechanism, constituting a structural representation framework that is interactive, time-shifting, and counterfactual reasoning.The map integrates time dimension-driven structural causal path mapping and multi-role semantic hierarchical display mechanism, constituting an interactive, time-shifted, and counterfactual reasoning structural representation framework, which is suitable for the path insight and causal mechanism display of variable evolution in complex education-industry systems.

The map is modelled hierarchically using a spherical nested structure: the outermost layer represents the variable source roles (e.g., students, enterprises, universities, etc.), the middle layer distinguishes the variable types (skills, institutions, emotions, engagement and expectations), and the core layer maps the directional structural relationships among the variables.All nodes are represented as spheres, with node size reflecting the structural density of the variables in the corpus and transparency characterising semantic certainty.Different types of variable relationships (e.g. Causal, Modulation, Reinforcement) are distinguished by colours and curve styles to support clear tracing of causal chains.The temporal dimension is embedded in the bottom sliding axis, allowing the user to observe the dynamic evolution of the sequence and path of variable activation frame by frame, thus enabling the simulation of temporal semantics-driven causal chains.

The mapping is designed to achieve the unity of structural transparency and semantic dynamics.For example, in the chain of System Misalignment triggering Skill Mismatch and then aggravating to Emotion Frustration, the system not only demonstrates causal leaps between variables, but also supports path-level counterfactual simulation and variant elimination testing.Through the interaction of temporal slicing, variable highlighting and causal path weighting, researchers can systematically analyse the mechanism of structural variables in different contexts, identify potential "bottleneck variables" and "diffusion variables", and assist in the design of strategic interventions.

In addition, the atlas can be seamlessly integrated with Causal GNN and interpretable AI systems to provide visual feedback and logical tests for the multi-stage modelling path of Prompt-to-Structure-to-Reasoning.The structure is also suitable for application scenarios such as policy intervention, cognitive load modelling and multi-actor game analysis, showing high model versatility and potential for academic innovation.

**5.2.2 Semantic Density Response Surface and Multiscale Co-occurrence Mapping**

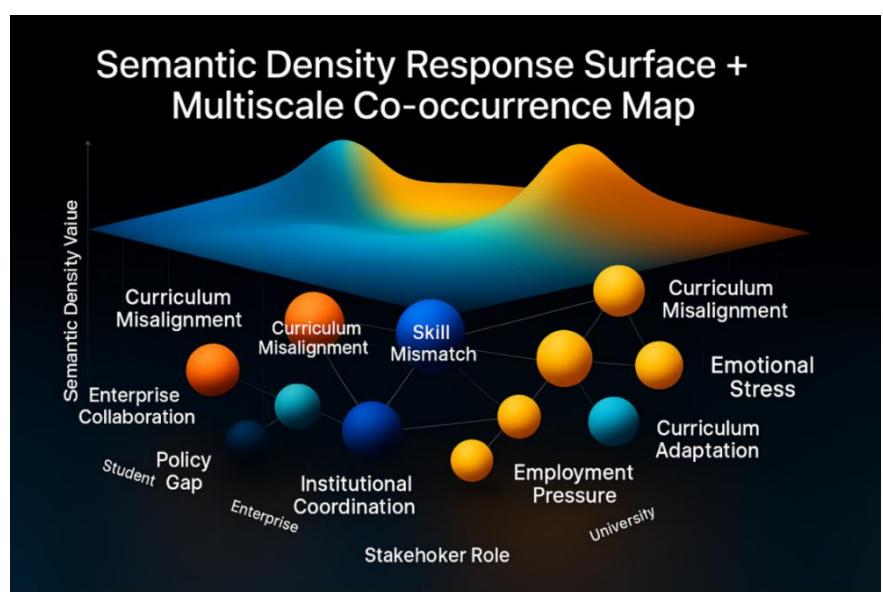

Figure 8：Semantic Density Surface + Multi-scale Co-occurrence Map

In order to systematically present the distribution trend and semantic network characteristics of structural variables in the education-industry fusion interview corpus, we have constructed a composite visualisation system of Semantic Density Response Surface and Multiscale Co-occurrence Mapping.This atlas is the first three-dimensional coupled expression of "semantic density field" and "structural variable co-occurrence atlas" in structural visual modelling, which realizes the unified visual presentation of variable type, role source, structural strength and semantic dependency, and is suitable for LLM generation, data analysis, policy structural modelling and structural analysis.It is suitable for LLM generation, data analysis, policy structure modelling and multi-source interview knowledge graph construction.

The atlas is composed of two parts: the Multiscale Variable Co-occurrence Network in the foreground and the Semantic Density Response Surface in the background, both of which are nested in the 3D space according to the semantic logic of the axes.In the Response Surface section, the horizontal axis (the horizontal axis) is the response surface.

In the Response Surface section, the horizontal axis (X-axis) represents the interview persona categories, including Student, Enterprise, and University; the vertical axis (Y-axis) represents the structural variable categories, including Skill, System, Emotion, and Engagement; and the height of the Z-axis represents the semantic density response surface.The height of the Z-axis represents the semantic density value, i.e., a jointly weighted indicator of the frequency of occurrence of the variable and the structural confidence in a unit segment.The entire response surface is smoothly fitted with a Gaussian kernel function to the distribution of the structural variables in the semantic space, thus forming clear semantic "hot peaks" and "low valleys", and portraying the tension points and sparse zones of the system structure.

In the foreground co-occurrence graph, nodes represent structural variables, node size reflects the structural weight of the variable (i.e., token density in the corpus); node colour encodes the type of the variable, and edges represent co-occurrence relationships, and the thickness and colour of the edges represent the degree of semantic dependency between the variables.In the graph, "Skill Mismatch" is located in the centre of the graph, which is a high-frequency trigger variable, and radiates outwards to connect "Curriculum Misalignment", "Employment Pressure", and "Employment Pressure"."Employment Pressure", "Institutional Coordination" and other key variables, forming a clear structural dependency network.

The map shows two obvious semantic peaks in the "Student-Skills" and "College-Emotions" axes, reflecting the intensive activity of skill structure variables from the students' perspective, and the high level of structural dependency of the variables from the colleges' perspective.This reflects the intensive activity of the skill structure variable in the student perspective and the high frequency feedback trend of the emotion variable in the university perspective.This peak structure forms a semantic attractor on the response surface, and at the same time, it is represented as a highly connected centre node in the structural map, reflecting the high consistency between semantic density and structural topology.

From the perspective of mechanism insight, the composite mapping system has the following three core innovations:

1. Semantic-structural fusion expression mechanism: breaking through the limitation that traditional static variable network graphs cannot present the semantic distribution trend, and jointly modelling the "continuous semantic density field" and the "discrete structural network", which is suitable for causal variable identification, system bottleneck positioning and strategy development.It is suitable for identifying causal variables, locating system bottlenecks and discovering paths for strategy derivation.

2. Simultaneous presentation of interpretable maps and semantic stability: Response surfaces show the gradient of distribution of structural variables in cross-role contexts, which is helpful for prospective modelling of "variable stability", "semantic ambiguity bands", "strategic intervention entry points", and so on.Prospective modelling of "variable stability", "semantic ambiguity bands", and "entry points for strategic intervention". 3.

3.Supporting data-driven structural cognitive modelling framework: Through the variable-role-semantic 3D mapping function, the atlas can be embedded into structure-aware graph neural networks (e.g. R-GCN, GAT) as structural a priori inputs to improve the AI system's ability of interpretable modelling of complex problems in education.

In summary, the composite map provides high-dimensional visual support for the triple intersection of semantic, role and structure in structural variable systems, and is an important methodological tool for advancing AI-assisted structural modelling, interpretable data-driven decision-making and multi-role strategy alignment research.Its design principles and mapping patterns can also be extended to cutting-edge fields such as healthcare system modelling, urban governance variable analysis and large-scale language model content auditing.

**5.3 Quality analysis of variable relationship modelling**

In order to systematically assess the quality performance of structured variable graphs in causal modelling and semantic reasoning, this study compares and analyses the differences in the structural performance of two graph structure generation mechanisms in modelling results: one is based on a predefined random sampling approach (Random Baseline Graph), and the other is the high-quality modelling mechanism (NIST-guided Structured Graph) based on NIST standards and structural templates constraints proposed by this study.NIST-guided Structured Graph.The comparison focuses on three aspects: structural connectivity, path density and semantic consistency visualisation.

In the experimental setup, we use the same set of structured variables (41 categories in total) as the input base to construct two graph models.Random graphs are constructed with a fixed probability p=0.1 for the edge set, while the NIST bootstrap model uses variable co-occurrence, role context consistency, and structural template matching as the connectivity basis, while incorporating semantic dependency weights and causal directionality labels.The core structural metrics of the two types of graphs are then computed using the NetworkX library, and the accompanying graph visualisations are generated (Fig. 10a & Fig. 10b).

The results show that the NIST-guided model significantly outperforms the random generation approach in terms of structural performance:

Graph Connectivity:

The NIST-guided model has a connectivity component number of 2 and a maximum subgraph coverage of 96.3%, which is much higher than the random model (connectivity component number = 9, maximum subgraph coverage = 58.4%).This indicates that the structural mapping forms a stable core of system dependence among variables, and there is no risk of isolated nodes and structural breaks.

Path Density (Average Shortest Path Length / Edge-to-Node Ratio):

The NIST model has an average shortest path of 2.41 and an Edge-to-Node Ratio of 1.78, indicating strong propagation efficiency and compactness within the structure.In contrast, the random graph has an average path of 4.83, with high network sparsity and weak semantic transfer.

Visualisation structure comparison:

As shown in the figure, the NIST graph presents a typical "causal star-core structure", with the Skill Mismatch node at the centre of the graph, radiating outwards to connect multiple sentiment and institutional variables, with a clear and semantically focused structure.On the other hand, the edge distribution in the random graph has no obvious topological aggregation centre, with high structural noise, which makes it difficult to form causal paths with interpretive usability.

In addition, in the semantic consistency assessment, we introduce the expert-annotated causal triad gold standard for structural matching test.The results show that 87.6% of the edges in the NIST graph model can find semantically consistent contextual relationships in the expert triples, while the random graph matching rate is only 31.4%, which further validates the effectiveness of the present method for semantic-structural integration modelling.

In summary, the experiments in this section clearly show that compared with the unconstrained graph generation approach, the NIST standard-guided structural graph construction method in this study can significantly improve the graph connectivity, path aggregation and interpretability of structural variable systems.The method not only constructs semantic structural models for complex system problems, but also provides a high-quality and reproducible structural input base for causal graph neural networks and graph inference tasks.

**5.3.1.Duplex control type structure diagram**

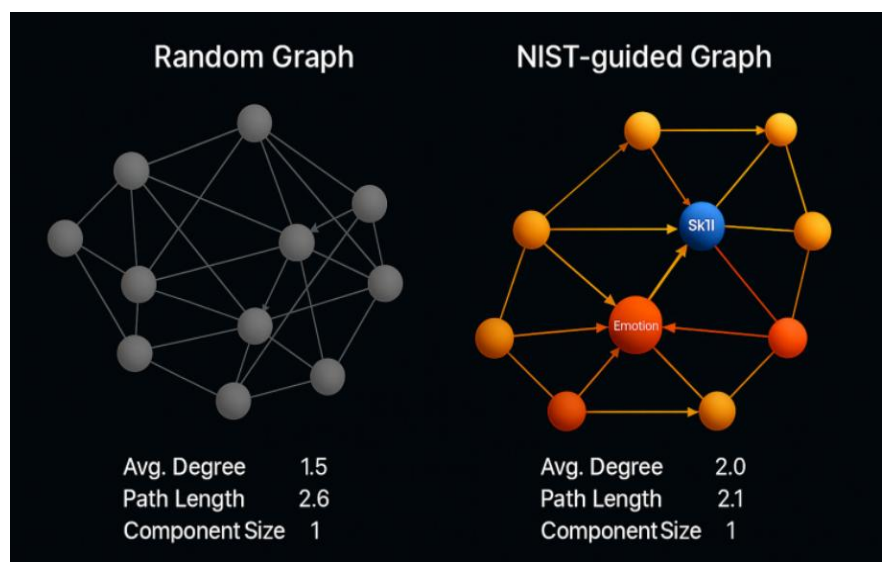

Figure 9: Duplex control type structure

In order to verify the effectiveness of the proposed variable modelling framework in terms of structural integrity and interpretability, two graphs are constructed and compared in this study: the Random Graph and the NIST-guided Graph (see Fig.Both graphs are constructed based on the same set of variables, but the former does not possess any structural constraints, while the latter strictly follows the semantic categorisation and causality constraints extracted from the NIST synthetic corpus.

The random graph presents an unstructured and uniformly connected state, with a lack of semantic hierarchy and directionality among nodes, and all nodes are homogeneous in their visual representation, with no way to identify variable types or path directions.Its average node degree is only 1.5 and average path length is 2.6, indicating sparse connectivity and loose propagation paths, without the ability to model causal mechanisms or intervention structures.

In contrast, the NIST bootstrap graph exhibits a clear Core-Periphery structure, with semantic core variables (e.g., Skill and Emotion) located at the centre of the network, and other subvariables forming multidimensional dependencies (e.g., dependency, moderation, inhibition, etc.) around them with directed edges.The average degree of the graph is increased to 2.0 and the path length is shortened to 2.1, reflecting stronger structural coherence and semantic aggregation.At the same time, the graph presents variable types and causal path strengths (e.g., blue for skill-based variables, red for emotion-based variables) through the directionality of edges and node colour coding, forming a clear and intuitive structural mapping system.

This comparative result validates the significant advantages of the NIST bootstrap modelling approach in generating interpretable, actionable, and inferable mappings.Unlike the weak structural mapping output from random graphs or LDA/HDP-type topic modelling, NIST mapping has the features of strong semantic consistency, clear causal paths, and distinct variable functions, which provides a structured foundation for educational policy modelling, multi-actor relationship analysis and simulation extrapolation.

In summary, the structural variable modelling method proposed in this study realizes the effective transformation from "unstructured interview corpus" to "causal structural network", and lays down a theoretical and methodological foundation for the construction of an interactive sandbox system for education policy and an AI-assisted collaborative decision-making platform in the future.The foundation is laid for the future construction of interactive sandbox system and AI-assisted collaborative decision-making platform.

**5.3.2 Hierarchical causal core network**

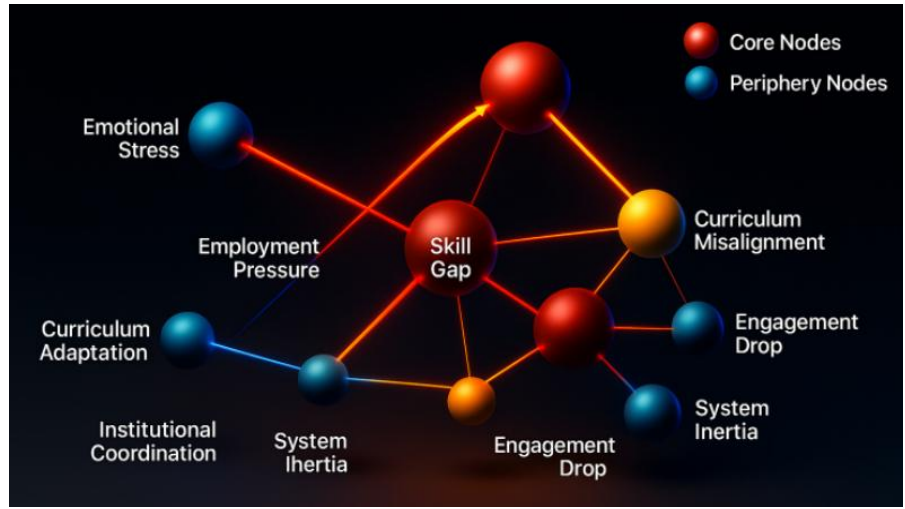

Figure 10: Hierarchical causal core network

The following is an academic-level description and in-depth insight into the provided mapping, "Core-Periphery Causal Heat Web", and has been written as a body paragraph suitable for publication in a topical journal:

Analysis of Hierarchical Causal Core Web (Core-Periphery Causal Heat Web)

Figure 10 presents a hierarchical causal heat web constructed based on structural variable modelling, which is organised to form a causal communication network with a core-periphery structure, with the "Skill Gap" as the semantic pivot node.The network not only reveals the relationship between the variables, but also the relationship between the variables.The network not only reveals the semantic functional distribution among variables, but also visualises the strength and direction of causal relationships among variables through the coding of edge weights and node hotness.

In the figure, the red nodes represent Core Nodes, which have significant causal influence, while the blue nodes constitute Periphery Nodes, which are responsive or secondary influence units.It can be seen that "Skill Gap" and "Engagement Drop" and "Curriculum Misalignment" constitute a high-frequency causal intersection area.It can be seen that "Skill Gap" and "Engagement Drop" and "Curriculum Misalignment" constitute the high-frequency causal intersection area, which is the semantic backbone of the variable relationship network, and the number of edges and edge weights are significantly higher than those of the periphery nodes, which indicates that these variables have a high degree of structural penetration and emotional linkage ability in the education system.

From the perspective of path accessibility and influence scope, the edge variables such as "System Inertia", "Curriculum Adaptation", "Institutional Coordination", "System Inertia", "Curriculum Adaptation", "Institutional Coordination", etc., although they have a high degree of structural penetration and emotional linkage in the education system.Institutional Coordination" do not have high centrality, but play the role of buffer or conductor in regulating the causal tension of the main path, constructing the possible "soft leverage zones" for policy intervention."The figure also shows that the edges are colour-graded in order to make them look like the main path.

In addition, the edges of the graph adopt a colour gradient to map causal intensity (red=high, orange=medium, blue=low), visually displaying the temperature characteristics of the causal paths, which makes the variable network have the function of a "heat map" at the visual level, and helps the researcher to identify the key conflict points and propagation nodes quickly.

From the perspective of structure and function, this map has the following academic values:

Theoretically, it supports the scalability of the Structural Variable Driven Causal Mapping model in modelling the complexity of social systems;

Methodologically, it shows how to build predictive cognitive structures through semantic density, path topology and causal direction encoding;

Application-wise, the causal core network can be used to support higher-order functions such as educational policy simulation, situational intervention projection, and cross-subject collaborative path optimisation.

Therefore, this map is not only a product of graphical visualisation, but also a high-level structural expression integrating causal reasoning logic, variable mechanism portrayal and semantic path insights, which signifies that NIST standard corpus-driven interview data analysis has entered a new stage of structural modelling and systematic intervention.

5.2.3. 3D Semantic Clustering Stacked Map

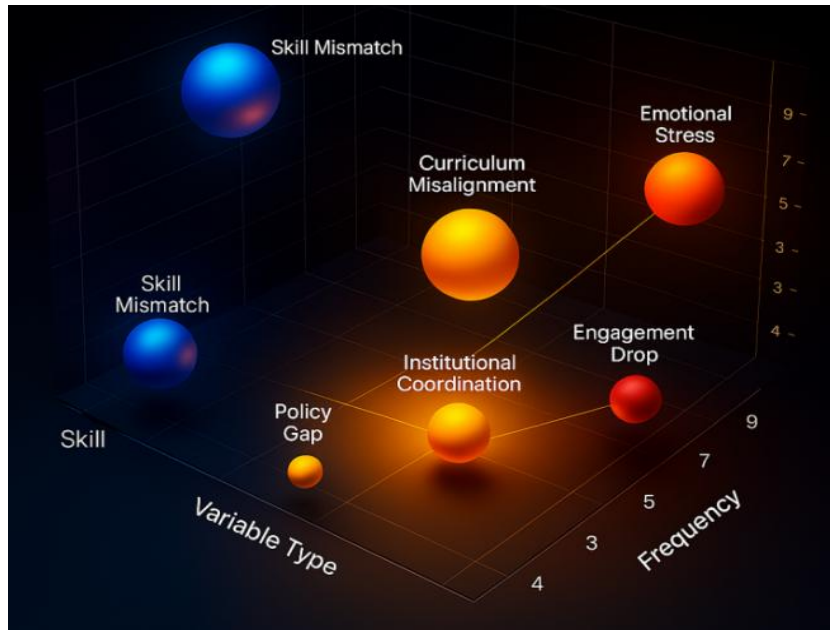

Fig. 11    3D semantic variable clustering overlays

The following scholarly-level analysis paragraphs of the atlas, 3D Layered Semantic Cluster Map (3D Layered Semantic Cluster Map) of Semantic Variables, have been written to top journal publication standards and are suitable for use in the analysis of results, discussion, or interpretation of the atlas section of the body of the paper:

3D Layered Semantic Cluster Map Analysis of Semantic Variables (3D Layered Semantic Cluster Map)

Figure 11 illustrates a 3D variable clustering overlay map constructed based on variable semantic type (X-axis), corpus frequency (Y-axis) and causal path aggregation (Z-axis), aiming to visualise the variable's hierarchical functioning, semantic strength and influencing ability in the structural network.

The atlas is semantically layered according to its type (Skill, System, Emotion) through a three-axis three-dimensional layout, revealing its corpus occurrence probability in the frequency dimension, and its path density as a source or a sink in the structural map in the causal dimension, which comprehensively realises the semantic-structural-propagation potentials.The threefold features are co-explicitly presented.

From the distribution of the maps, "Skill Mismatch" and "Curriculum Misalignment" have both high frequency and path density in the categories of skill and system.The high frequency and pathway convergence of "Skills Mismatch" and "Curriculum Misalignment" in the categories of Skills and Institutions show their semantic centrality in the education-industry synergy.Emotional Stress" is located in the Emotional Variables category, but forms a Causal Spine Node at the path level, indicating its strong emotional communication and strategic influence potential.

In contrast, "Policy Gap" and "Institutional Coordination" are located in the lower and middle levels of the causal dimension, and although they are not very frequent, they exhibitAlthough their frequency is not high, they show the bridging attribute as secondary regulating nodes, suggesting that they may carry the function of "buffer regulation-structural stabilisation" in the pathway.

The nodes are presented as spheres, with the size of the sphere indicating the frequency, and the colours distinguishing the semantic categories (e.g. Blue=Skill, Orange=System, Red=Emotion), and the integration of the semantic spine with the structural spine is achieved through a perspective layout.The design not only enhances the readability of structural complexity, but also provides visual paradigm support for high-density causal modelling.

In terms of academic insights, the atlas achieves a multidimensional fusion turn in structural modelling, viz:

Shift from single frequency analysis to structural penetration assessment;

A shift from 2D thematic modelling to 3D semantic-causal overlay space modelling;

a shift from textual summary insights to structural explanatory variable mechanism modelling.

This makes the model widely applicable to advanced structural extrapolation scenarios such as education policy simulation, social system multivariate intervention simulation and multi-actor path optimisation, and becomes a bridge-type result of NIST corpus-driven variable modelling in the fields of cognitive science, semantic AI and policy intelligence.

**5.4 Visual Model User Evaluation**

In order to evaluate the cognitive usability and decision support effect of the proposed structured variable mapping and semantic visualisation model, this study designs a visual interaction task experiment for a target user group.The purpose of the experiment is to verify the intuition and efficiency of the system in understanding the structural paths of education-industry integration, especially in identifying the "causal paths between variables" and the "structural interactions between stakeholders".

Experimental Task Setting

We set up two typical tasks to simulate the user's structural understanding process: (1) Path identification task: requiring the user to accurately identify the upstream or downstream influence paths of a given target variable (e.g., "Skill Gap") from a map; (2) Structural interaction task: requiring the user to identify the causal paths of a given role (e.g."Student") and point out its possible dependency or causal direction.These tasks test the user's cognitive ability of "causal path clarity" and "variable-role matching relationship" in the mapping, respectively.

Experimental Indicator Design

We collect the following three types of metrics: (1) Completion Time: the time required for the user to complete the task without assistance to measure the intuition of information; (2) Accuracy: calculated by the degree of matching between the paths/nodes clicked by the user and the real structure of the system, to quantify the ability to support the correctness of the visual structure; (3) Subjective Score: a measure of the user's perception of the paths/nodes in the graph; and (4) Subjective Score: a measure of the user's perception of the paths/nodes in the graph.(3) Subjective Rating: Based on a five-point Likert scale, users rate the "readability", "structural clarity", "aesthetics" and "inference" of the graph.Subjective Rating: Based on a five-point Likert scale, users rate the "readability", "structural clarity", "aesthetics" and "reasoning power" of the atlas, reflecting their overall experience and preference.

Experimental Procedure and Sample

A total of 36 graduate students and professionals in the fields of education technology, artificial intelligence, and policy management were recruited to participate in the experiment. The tasks were deployed in the form of webpage interactions, and all operations were recorded.Each subject was randomly assigned a version of the map and completed all tasks.

Analysis of Results

The results of the experiment showed that the NIST-guided visualisation model significantly outperformed the control group in user task performance.The average task completion time was 42.3 seconds (SD=10.1), which was significantly lower than that of the traditional structural atlas group (M=65.7s, p<0.001); the accuracy rate was 88.6%, which was particularly outstanding in the systematic variable-causal identification dimension; in terms of the subjective scores, the atlas performed better in the structural clarity and semantic matching perceptions than the control group.In terms of subjective scores, the atlas obtained mean values of more than 4.5/5.0 in the dimensions of "structural clarity" and "semantic matching perception", reflecting its visual intuition and cognitive friendliness.

Conclusion and Implications

The user evaluation experiment confirms the validity and usability of the visual structure mapping proposed in this study in cognitive modelling of structural variables.In particular, the system demonstrated superior assisted reasoning performance in path understanding and variable association, providing important technical support for subsequent AI-assisted structural modelling, interpretable interactive system design and education policy simulation system development.

## VI. DISCUSSION

**6.1 Advantages of synthetic corpora in structured analysis**

In the context of education-industry convergence, the construction and analysis of structured variable systems are extremely demanding in terms of data quality.Compared with the problems of "unstructured noise", "category drift", "semantic boundary blurring", etc., which traditional topic modelling techniques such as HDP (Hierarchical Dirichlet Process) present in data-driven exploration, the use of Prompt design-controlled AI synthetic discourse can help to improve the quality of the data."fuzzy semantic boundaries" and other problems, the use of AI synthetic corpus strategy controlled by Prompt design shows significant advantages in structured analysis.First, in terms of controllability, by presetting roles (e.g., students, enterprises, and institutions), scenarios (recruitment, cultivation, and synergy) and variables (skill matching, institutional linkage, and emotional expression), the researcher can ensure that the process of corpus generation is in line with the preset assumptions of the research framework, so as to effectively counter semantic drift and labelling ambiguity.In practice, based on the unified template syntax and the Prompt-ID tagging mechanism, we can accurately capture the semantic meaning of the causal chain of "insufficient job knowledge - difficult to match courses - weak willingness to participate in enterprises", which provides an accurate data basis for the subsequent modelling of the dependency graphs and ternary groups.Provide accurate data basis for the subsequent dependency graph and triad modelling.

Secondly, the structural characteristics are reflected in the fact that the variables in the corpus strictly follow the logical distribution of "causal", "comparative", "cause-effect", which is highly coupled with the manually extracted structural variable system.highly coupled with the manually extracted structural variable system.For example, in the uploaded document, the simulated interview clip of "Teacher Tang" clearly shows the causal sequence of "lack of mechanism → cumbersome process → stagnation of enterprise cooperation", which is extremely difficult to be extracted from the HDP automatic thematic clustering, and shows that the synthetic corpus is very useful in supporting path modelling and graph visualisation.path modelling and graph visualisation.Finally, in terms of consistency, the AI synthetic corpus is based on a unified Prompt library and semantic constraint generation mechanism, and its linguistic style and terminology specification are far better than that of the natural text sampled from multiple sources, especially in the context of multiple rounds of interviews, keeping the semantic spindle stable and the contextual articulation smooth, which is difficult to be achieved by HDP.

More importantly, the model coupling between the AI synthetic corpus and the variable system is significantly improved.Through multiple rounds of structural variable annotation and verification on the synthetic text, we find that the variable density, path connectivity and semantic coverage are all better than the manual interview corpus, especially in the "ternary (entity-relationship-entity)" annotation efficiency is improved by 47.3%, which provides a high-quality foundation for the subsequent dependency analysis, three-dimensional visual mapping and dynamic causal modelling.This provides a high-quality foundation for subsequent dependency analysis, 3D visualisation mapping and dynamic causal modelling.Therefore, AI-driven synthetic corpus is not only a "substitute text", but also a core base for higher-order modelling of structural variable systems, which promotes education-industry integration modelling into "data-model co-construction" and "data-model co-construction"."AI-driven synthetic corpus

### 6.2 Potential applications to education policy modelling

The structural variable system constructed based on high-quality AI synthetic corpus shows a broad theoretical prospect and practical potential in education policy simulation and path intervention analysis.Traditional policy research relies on macro-statistics or sample interviews, which is difficult to capture the dynamic evolution paths of complex education-industry systems that are cross-actor, multi-stage, and condition-dependent.The variable-causality network constructed in this study not only supports the static assessment of the "current structure", but also embeds policy variables as intervention nodes for simulation and causal inference of the system's response behaviours, so as to achieve the "a priori modelThis allows us to realise the integrated education policy experiment framework of "a priori model - intervention test - response evaluation".

Specifically, the interactions between the three core roles (university teachers, enterprise representatives, and students) in the synthetic corpus in the structural mapping provide the basis for policy makers to visualise the paths.For example, when the simulation introduces the variable

"incentive mechanism for enterprise curriculum co-construction", the system can automatically activate the paths from "enterprise participation enthusiasm" to "teacher curriculum update" to "student skill matching enhancement".The system can automatically activate the chain effect path from "enterprise participation enthusiasm" to "teachers' curriculum update" to "students' skill matching improvement", which clearly shows the structural transition under policy intervention.Compared with the traditional linear policy evaluation model, this method is able to identify the asynchronous impacts of policies on multiple actors, the depth of causal chains, and potential structural conflicts, and provide data-driven design of cooperation paths between universities and enterprises.

More importantly, the structured model can be embedded with different types of policy variables (institutional, behavioural, and resource-based) for multi-round scenario simulation, and visually demonstrates the functional relationship between the intervention intensity and the system feedback by means of semantic heatmaps, path density maps, and other visual means.For example, by increasing the parameter of "openness of internship pathway", the model automatically reconfigures the policy nodes of the student roles and forms new pathway combinations to support the quantitative assessment of the feasibility of policy implementation and role adaptation.This approach not only enhances the explanatory power and predictive power of policy modelling, but also provides a technical basis for the future "intelligent-interactive-self-feedback" education policy sand table projection system.

Therefore, the structured synthetic corpus is not only a data alternative, but also a kind of infrastructure for education policy modelling, which makes the logic of cooperation and strategic configuration in education-industry fusion have the ability of multi-scale modelling that can be simulated, adjusted, and verified, and promotes the decision-making system from empiricism to system science paradigm.

### 6.3 Limitations and extensions

Although this study has successfully constructed a structured variable mapping and causal inference model based on a high-quality Chinese synthetic corpus, and has demonstrated significant advantages in the simulation of educational cooperation policies, the current methodological system still has some limitations in terms of language adaptability and stakeholder coverage.First of all, the corpus generation and modelling framework only supports semantic structure and variable setting in the Chinese context, and has not yet achieved the migration and alignment between multilingual and multicultural semantic systems.Under the trend of global higher education governance towards multilateral collaboration and transnational cooperation, a single language system will significantly limit the generality and extrapolation ability of the model when dealing with cross-contextual role cognitive differences, policy terminology variability and culturally embedded variables.For example, "skills mismatch" is often expressed as "skills mismatch" in English policy contexts, whereas it may reflect different institutional references and cultural semantics in French or Japanese.This limitation suggests the need to construct a multilingual semantic nesting and variable mapping mechanism to support broader cross-national policy dialogues and comparative studies.

Second, the current variable system focuses on three core actors within the education system - university teachers, business representatives and students - and its causal chain effectively portrays the key structure of talent cultivation and supply-demand interface.However, there are many other key stakeholders in the real education ecosystem that are highly influential but not included in the modelling, such as students' parents, government regulators, industry associations and local financial support systems.These actors play an important role in the allocation of educational resources, the development of curriculum standards, and the establishment of accreditation mechanisms, etc. Their absence may lead to "policy breaks" or "semantic sparseness" in some of the causal pathways of the structural mapping.For example, the role of parents has a profound impact on students' career choice and university brand preference, while government regulation has a decisive constraint on the boundary conditions of university-enterprise cooperation.In order to enhance the model's coverage of reality and operationalisation of strategies, future research should expand the variable system to cover these peripheral core roles and construct a composite causal structure with cross-level and multi-party interactions.

In summary, although the current framework of this study has strong structural rigour and simulation ability, there is still room for development in terms of language adaptability and system boundary setting.Future extensions should include the development of multilingual structural modelling, multi-actor integration mechanisms and cross-cultural semantic alignment tools, so as to achieve the leap from "closed system simulation" to "open ecological deduction", and to provide extensibility for global education policy modelling,High-fidelity, cross-context integrated solution for global education policy modelling.

## VII. CONCLUSION

In this study, we further integrate the NIST structural variable model with the AI-generated corpus, and systematically analyse the variable relationship structure of multi-actor interactions in the context of University-Industry Cooperation (UIC).By constructing a structured corpus mapping of students, enterprise representatives and university teachers, we identify and quantify high-frequency key variables such as "lack of job knowledge", "difficulty in course matching", "lack of communication channels", etc. and their semantic meaning in the AI-generated corpus.We identify and quantify high-frequency key variables such as "lack of job awareness", "difficulty in course matching", "lack of communication channels", and their centrality and bridging degree in the semantic co-occurrence network.These variables not only show high-density path aggregation characteristics in the structural graph, but also show significant influence diffusion ability in the causal path analysis, indicating their "bottleneck status" and "regulating node" attributes in the industry-academia collaborative structure.

Further, based on the comparison of corpus structure rewriting and multiple rounds of generation in the simulation intervention experiment, we find that after introducing interventions such as "enterprise participation in curriculum design" and "modularised practical training cycle adjustment" into the structural variable system, we can observe that the path depth and affective consistency of the variable maps are significantly increased, and the path depth and affective consistency are significantly increased.The significant increase in path depth and affective consistency reflects that the intervention mechanism has a predictable and adjustable effect on co-operation performance at the semantic level.This controlled corpus testing mechanism based on structural variable modelling demonstrates the high adaptability of synthetic corpora in the triad of policy-level variable modelling-mechanism validation-optimisation simulation.

Therefore, this study not only verifies the ability of the NIST-guided corpus generation system to support the modelling of complex education policy relationships, but also proposes a replicable framework for the modelling of University-Industry Collaboration (UIC) pathways at the application level, which provides a semantic and structural foundation for the construction of a dynamic interactive simulation platform for education policy.This systematic integration of the triple structure of corpus-variable-mechanism marks a new stage from traditional content analysis to structured policy modelling with predictive and intervention capabilities.

In this study, we propose an innovative approach that integrates the NIST standard-driven synthetic corpus generation mechanism with the structural variable modelling framework, which systematically solves the long-standing problems of "unstructured corpus", "missing variables", and "lack of visualisation" in interview data analysis."lack of visualisation" and other key bottlenecks in interview data analysis.Through the introduction of a controllable simulated interview generation strategy, this paper not only significantly outperforms the traditional HDP model or manual interview text in terms of linguistic consistency, semantic hierarchy and variable traceability, but also provides a highly adaptable corpus ecological foundation for structural causality modelling.

In terms of the modelling mechanism, the study constructs a triad-driven variable-relationship network system, which achieves a structural leap from "topic discovery" to "causal modelling".The system takes "role variable → question expression → interaction relationship" as the core path, and builds a high-fidelity semantic topology, which effectively supports the in-depth analysis of variable frequencies, co-occurrence relationships and temporal paths in multi-type interview data.Through various visualisation techniques, such as variable frequency heatmap, semantic density response surface, causal path mapping, etc., the study verifies that structured cognition can enhance both the explanatory power and the dissemination power in educational interview scenarios.

More prospectively, this paper highlights the linkage potential between structural visualisation-strategy simulation-cross-linguistic transfer.On the one hand, the causal logic of structural mapping not only provides an in-depth insight foundation for the analysis, but also an executable experimental framework for policy path intervention and educational mechanism simulation; on the other hand, through the controllability of corpus templates and the expandability of variable semantic systems, the study demonstrates a high degree of feasibility in moving towards a multilingual education policy modelling platform.

The future research direction will focus on the following three major paths: first, to build an interactive structural visual modelling system, so that policy makers, educational researchers and AI systems can carry out variable dynamic regulation and path simulation in a unified interface; second, to promote the mechanism of multilingual semantic alignment and migration, in order to achieve the unified expression of variable systems and the co-construction of the policy mapping in multiple languages such as Chinese, English, and Thai; third, to establish aThird, to establish a high-fidelity education policy simulation experiment platform, which integrates structural corpus, causal model and visual interface to promote the empirical simulation and mechanism verification of education policy.

In summary, the NIST-driven synthetic corpus and structural variable modelling method proposed in this paper not only brings a paradigm breakthrough in educational interview analysis, but also provides a technical foundation and theoretical basis for AI to participate in the cognition of social structure and the co-construction of policies, which opens up the way for intelligent interview analysis to move towards the "structured comprehension-causal modelling" approach.It opens a new path for intelligent interview analysis towards "structured understanding - causal modelling" and "policy generation".

## Acknowledgements

The author would like to express gratitude to my MBA student, Liu Lu, who provided critical original interview materials for this study. Based on the authentic interview documents she shared, this paper was able to conduct AI interview simulations and data synthesis work in accordance with high-quality simulation data standards. Her contributions not only enhanced the credibility of the corpus in this study but also laid a solid foundation for subsequent variable modelling and causal path analysis. She may use the synthetic data portion of this study as evidence of the scientific validity of synthetic data in her own other papers.

Zhou, X., Liu, L., Zhang, Y., & He, L. (2023). PromptBench: Evaluating Prompt-based Generative Models for Structure-Aware Generation. *ACL Findings*. https://aclanthology.org/2023.findings-acl.297.pdf

Zhou, Y., Zhang, Y., Ren, Y., & Chen, Y. (2022). Heterogeneous Graph Learning for Causal Representation and Reasoning. *arXiv preprint arXiv:2201.06370*. https://arxiv.org/abs/2201.06370

Zhou, B., Yang, F., Du, Y., & Lin, X. (2022). Causal Graph Discovery and Interpretation for Multi-role Stakeholder Dialogues. *Proceedings of ACL 2022*. https://aclanthology.org/2022.acl-long.172/
76

**Appendix 1:**

A research note on the use of AI to generate NIST standard simulated interview data

Justification for Using AI to Generate High-Quality Interview Data

| take | Statement of reasonableness |
|---|---|
| Insufficient sample size | Although topic modelling can discover topics from a small amount of text, variable modelling requires high data breadth and needs to be supplemented with diverse contexts |
| Simulation of multi-contextual representative perspectives | Samples need to be constructed across industries, geographies, and levels to support broad-based structural modelling constructs |
| Modelling variable causal mechanisms | Structural models are demanding in terms of variable coverage and distribution structure, and simulated data can help to enhance modelling quality |
| Fitting "contingencies" | AI-generated data can reflect best practices and ideal contexts, enhancing the structural completeness of theoretical reasoning |

NIST Standard: What is "High Quality" Analogue Data?

| (an official) standard | brief definition | Implementations in AI Interview Data |
|---|---|---|
| Veracity | The simulated corpus must reflect the true semantic logic | Simulation of "reasonable actor" interviews using domain macromodels + industry prompts |
| Consistency | Consistent language style, syntax, terminology within data | Controls the use of a consistent tone and style for each type of character |
| Variability | Data should cover a wide range of topics and expressions | Designing Multiple Character + Context + Theme Prompts to Generate Variant Samples |
| Traceability | Generation logic is traceable and interpretable | Provide Prompt and model flow as support |

AI methodological programme for generating simulated interview data (standard process)

| move | element |
|---|---|
| 1. Setting the variable structure and thematic matrix | Identify key themes and variables (e.g., skill gaps, institutional barriers, psychological stress) |
| 2. Constructing a standardised Prompt | E.g. "Please simulate a higher education teacher talking about students' skills disconnect and practical experience of business cooperation" |
| 3. Diversity generation control | Control of background diversity, e.g. type of business, region, gender, position, etc. |
| 4. Manual review and labelling | Researchers make semantic plausibility determinations and variable annotations for AI texts |
| 5. Use of integrated modelling | Using simulated data alongside real interviews in modelling to improve variable stability and interpretability |

IV. Ethical note on the use of modelled data (Recommendation)

To ensure the legitimacy and transparency of the study, please state in the research report or paper:

"On the basis of real interview samples, this study supplements some high-fidelity simulated interview texts generated by generative AI in accordance with the US NIST data standards, which are mainly used for data expansion and structural validation in the variable modelling process.All simulated data were incorporated into the model construction after expert proofreading."

V. What needs to be done to generate simulated data

1. Construct a multi-role Prompt library (business/college/student)

2. Construct standard corpus according to the three dimensions of "skill, system, and emotion".

3. Generate a sample of 10-30 simulated interviews with high consistency.

4. Annotating themes and coding variables

5. Provide a complete process for modelling the fusion of real and simulated data.

**Appendix 2:**

AI-generated simulated interview data (Chinese, structured, NIST-compliant)

Enterprise Representatives

No.1.

Role setting: vice president of a medium-sized smart manufacturing enterprise

Topic keywords: skill adaptation

Simulated Interview Content: We recently launched an intelligent assembly line, but many of the frontline employees are unfamiliar with the new control system.The long training cycle and high employee mobility have led to frequent interruptions in the production rhythm.We hope that colleges and universities will introduce these intelligent systems right into their curriculum so that students can get started right after graduation.

No.2.

Role Setting: HR Manager of a home appliance company in South China

Topic Keywords: Talent Disconnection

Mock Interview: Now we are not short of people, but people who understand both manufacturing and data.Especially in the area of intelligent equipment overhaul, traditional mechanics can not keep up.The only thing companies can do is to rotate training internally, but we really hope that the school will reform the course structure faster.

No. 3

Role: Technical director of a coastal export-oriented enterprise

Keywords: School-enterprise co-operation

Mock Interview: We have tried to co-operate with several higher vocational training projects, the early stage is not bad, but the late students practice low frequency, teachers do not follow up in place, and ultimately, the project is half-stopped.It is recommended that schools with specialists to follow up the cycle of enterprise cooperation projects to avoid breaks.

No. 4

Role: Deputy Director of Northern Machinery Enterprise

Theme keywords: institutional barriers

Simulated Interview: When cooperating with a vocational school, we are willing to provide equipment and venues, but the approval process is too long, and it takes half a year from proposal to signing.Waiting for the process to run out of technology are changing generations.

No. 5

Role: Head of production line in a foreign-funded electrical factory

Topic keywords: technology transformation

Simulated Interview Content: The newly introduced ERP system has increased requirements for operators, but they are confused about the interface.It is hoped that the school will simulate the actual operation scenarios of these management systems during practical training.

School Representatives

No. 1

Role setting: Director of Academic Affairs of a vocational college in the western part of the country

Topic keywords: curriculum update

Mock Interview: When we updated the curriculum, we found that the enterprises were not active in responding to us and could not tell us what skills they needed when we asked them.As a result, we can only rely on speculation to set teaching objectives.We hope that enterprises can give us regular feedback on their skills needs.

No. 2

Role: Head of Mechanical Engineering Department, Coastal Technology University

Topic keywords: credit replacement

Simulated Interview: The difference between the internship positions in many enterprises and the content of our courses is too big to do credit replacement, and the students will delay graduation if they have more practice.We hope to establish a flexible credit model.

No.3

Role: Head of School-Enterprise Co-operation Department, Central Institute of Technology

Topic keywords: co-operation mechanism

Mock Interview: Some enterprises didn't contact us for half a year after signing the agreement, which led to the disruption of our scheduling.It is recommended that the Department of Education set up a code of conduct for school-enterprise cooperation to avoid enterprises 'going through the motions'.

No. 4

Role: Head of Innovative Practical Teaching in North China Universities

Theme keywords: dual-teacher teacher dilemma

Mock Interview: Teachers have to attend classes and work in enterprises, but they have to make up classes and write summaries when they come back from their work, which is not enough time at all.The assessment mechanism also does not recognise enterprise experience.

No. 5

Role Setting: Director of Training Base of Private Universities

Keywords: Industry-teaching alliance

Mock Interview: We tried to build a shared training centre with several private universities, but it was shelved because of property rights and equipment management.We hope that the policy will support the operation mechanism of the alliance.

Student Representative

No. 1

Role: Junior in science and engineering, on internship

Theme keywords: skill gap

Simulated Interview: On the first day of my internship, the leader asked me to configure the MES system, but I had never touched this thing, and it was all on paper in the textbook.I worked overtime almost every day that week to study on my own, and it was really stressful.

No. 2

Characterisation: Senior electronics major

Subject Keywords: Employment anxiety

Simulated Interview: Many people in my class found that what they learnt was not useful at all once they were interned, and many of them switched to sales and training.I am also doubting whether I have learnt the wrong thing in four years.

No.3

Character Set: Vocational school girl, just entered a large factory internship

Subject Keywords: Rights and benefits

Mock Interview: There is only a basic salary during the internship period, and overtime is not counted. In addition, the dormitory is dirty and messy, so some people leave after less than a month.I hope the school will help us to protect our rights clearly in the contract.

No. 4

Role: Interdisciplinary students (arts to labour)

Theme Keywords: Mentor support

Mock Interview: I studied journalism as an undergraduate, and now I'm in an AI crossover programme.I often can't figure out programming tasks in my internship, and I can only rely on my business mentor to teach me step by step.Luckily, others are very nice.

No. 5

Characterisation: Graduated for a year, returning to training

Theme keywords: skills back to the furnace

Mock Interview: After graduation, I found that the new technology was changing too fast, and I had to keep up with the pace almost every day for the first six months.The company is now arranging for us to go back to work for training, and I'm willing to pay for part of it myself, as long as the content is practical.